\documentclass[aps,prb,draft,amsmath,twocolumn,floatfix,bm]{revtex4}

\usepackage{bm}
\usepackage{epsf}
\usepackage{array}
\newcommand\e{\varepsilon}
\newcommand\x{{\bf x}}
\newcommand\rbf{{\bf r}}
\newcommand\R{{\bf R}}
\newcommand\n{{\bf n}}
\newcommand\p{{\bf p}}
\newcommand\apq{a_{+}}
\newcommand\amq{a_{-}}
\newcommand\esc{\tau_{\rm esc}}
\newcommand\mls{\delta_1}
\newcommand\g{\hat {\cal G}}
\newcommand\N{{\cal N}}
\newcommand\tte{\tau_{\rm E}}
\newcommand\gesc{\gamma_{\rm esc}}

\begin{document}
%\draft
\title{Quantum Disorder and Quantum Chaos in Andreev Billiards} 
\author{M.G. Vavilov$^{a}$, and  A.I. Larkin$^{a,b}$ }
\address{$^a$Theoretical Physics Institute, University of Minnesota,
Minneapolis, MN 55455\\
$^b$Landau Institute for Theoretical Physics, Moscow, Russia 117940}
\date{\today}

\begin{abstract}
We investigate the crossover from the semiclassical to the quantum description of electron energy states in a chaotic metal grain connected to a superconductor. We consider the influence of scattering off point impurities (quantum disorder) and  of quantum diffraction (quantum chaos) on the electron density of states. We show that both the quantum disorder and the quantum chaos open a gap near the Fermi energy. The size of the gap is determined by the mean free time in disordered systems and by the Ehrenfest time in clean chaotic systems. Particularly, if both times become infinitely large, the density of states is gapless, and if either of these times becomes shorter than the electron escape time, the density of states is described by random matrix theory.  Using the Usadel equation, we also study the density of states in a grain connected to a superconductor by a diffusive contact.
\end{abstract}

\maketitle

%\newpage

\section{Introduction}
\label{sec:1}
Condensed matter physics usually deals with quantum disorder.
Quantum disordered systems contain random impurities with the
characteristic size smaller than the electron Fermi wavelength 
$\lambda_{\rm F}$. In these systems, electrons travel between scattering off impurities along classical trajectories, but the scattering is described by quantum mechanics. Averaging over impurity configurations
leads to quantum effects in disordered systems.\cite{WL}

Recently, new quantum systems - quantum dots -
were created.\cite{qd} In quantum dots the size of
irregularities is much larger than the electron wavelength, therefore electron motion may be determined by  a semiclassical theory.

In the semiclassical limit certain thermodynamic and kinetic quantities can be calculated by expansion in powers of the Plank's constant $\hbar$.\cite{LL}  This expansion corresponds to a series in powers of $\lambda_{\rm F}/L$,
where $L$ is the characteristic length scale of the system and 
$\lambda_{\rm F}$ is the electron Fermi wave length. 
For other quantities such semiclassical description is justified only for  times shorter than the Ehrenfest time $t_{\rm E}$. At longer times a crossover from semiclassical to quantum description of these quantities takes place. [Ehrenfest\cite{E} demonstrated that the semiclassical description of  motion of a quantum particle is justified for short times.] 

What is the Ehrenfest time in classically chaotic systems, where the classical motion is governed by exponential divergence of  trajectories?
It was shown in Refs.~[\onlinecite{LO,Z,AL1}], that the Ehrenfest time logarithmically depends on $\hbar$. Indeed, in semiclassical theory, a particle is represented by a wave packet  - a superposition of particle wave functions. The size of the wave packet  cannot be smaller than the electron wavelength $\lambda_{\rm F}$ and cannot be larger than the characteristic scale $L$ of variations of the potential energy. 
In chaotic systems the size of any wave packet exponentially increases in time. Thus, the wave packet increases from its initial size $\lambda_{\rm F}$ to size $d(t)=\lambda_{\rm F}\exp(\lambda t)$ as time $t$ increases and expands to scale of the system $L$ after the Ehrenfest time $t_{\rm E}=\lambda^{-1}\ln L/\lambda_{\rm F}$, where $\lambda$ is a typical value of the Lyapunov exponent.

We refer to a particle motion at time longer than the Ehrenfest time as a regime of quantum chaos. 
Quantum chaos affects different phenomena of condensed matter physics. In a metal with random long-range potential the weak localization correction to the a.c.-conductance at frequency $\omega$ is absent at high frequencies, 
$\omega t_{\rm E}\gg 1$.\cite{AL1} Nonetheless, at lower frequencies, 
$\omega t_{\rm E}\ll 1$, the weak localization correction to the
conductance reaches its universal quantum limit. 

The statistics of
electron levels is correctly described by the Gutzwiller trace
formula at large energies, $\varepsilon t_{\rm E}\gg 1$. At
small energies, $\varepsilon t_{\rm E}\ll  1$, energy levels obey the
Wigner--Dyson statistics. 
At intermediate energies, $\varepsilon t_{\rm E}\sim 1$, the weak
localization effects are important for understanding of the level statistics.\cite{AL2} 

The shot noise through a chaotic quantum dot is absent,  
if the electron escape time $\tau_{\rm esc}$ from the dot  is shorter than the Ehrenfest time, $\tau_{\rm esc}\ll t_{\rm E}$. In the opposite limit of long $\tau_{\rm esc}\gg t_{\rm E}$, the shot noise reaches its universal value.\cite{AAL,SS}

The semiclassical method in theory of superconductivity 
was developed by de Gennes and Tinkham \cite{dGT} and 
Shapoval \cite{Shap} for the description of the
diffusive electron scattering at surfaces. The applications\cite{dGT,Shap,dGbook} of the semiclassical method gave a basis for a belief\cite{GL,LOF} that in order to obtain electron properties of a superconductor, first, electron states in the superconductor can be found as solutions 
of semiclassical equations and, then, these solutions can be averaged over different classical trajectories.  However,  a more detailed analysis\cite{LO}  
demonstrated that in certain problems [\emph{e.g.} in calculations of the magnetic penetration depth as a function of an applied magnetic field], the semiclassical method does not work, and the averaging over classical trajectories gives a different result than that obtained by the standard technique of averaging over impurities.\cite{AGD} 

According to ref.~[\onlinecite{LO}], even for a system with smooth 
long-range  disorder, $L\gg \lambda_{\rm F}$, the semiclassical method is applicable only at time smaller than the Ehrenfest time $t_{\rm E}$. 
Paper~[\onlinecite{LO}] explains the discrepancy between semiclassical and quantum methods in theory of superconductivity: within the semiclassical approach electrons and holes move in opposite directions along identical trajectories inside a superconductor. Because quantum diffraction switches electrons and holes between different trajectories, such semiclassical description breaks down at times larger than the Ehrenfest time $t_{\rm E}$. 
The Ehrenfest time $t_{\rm E}$ and the Lyapunov exponent $\lambda$ were calculated for electron systems with smooth impurities of the size $L\gg \lambda_{\rm F}$ in ref.~[\onlinecite{LO}].

\begin{figure}
\centerline{\epsfxsize=4cm
\epsfbox{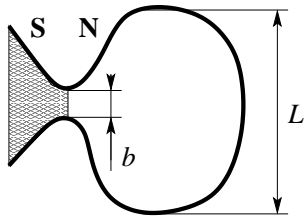}}
\caption {
A sketch of an Andreev billiard: a metal grain (N), connected to a superconductor (S). The size of the grain is $L$, and the size of the contact is $b$.  
}
\label{fig:1}
\end{figure}

In this paper we study the density of states in an Andreev 
billiard. An Andreev billiard is a small chaotic metal grain
or semiconductor quantum dot\cite{qd} connected to a superconductor through a (SN) contact. Electron reflection at the contact is described by the Andreev reflection mechanism,\cite{Arefl} when an electron is reflected as a hole, moving in the opposite direction. The density of states in Andreev billiards was studied both by
semiclassical\cite{LN,ScB,MBFB} and quantum methods.\cite{MBFB}
The semiclassical method gives an exponentially small but non-zero tail of the density of states for any finite energy, see refs.~[\onlinecite{LN,ScB}].
On the contrary, quantum calculations\cite{MBFB} within random matrix theory
predict a gap in the energy spectrum.\cite{fn1} [Both curves of the density of states are presented below in Sec.~\ref{sec:3.2.2}.] 

The difference between semiclassical and random matrix results for the density of states has attracted theoretical interest to this problem.\cite{TSA,AB,Bnew} 
Based on the approach of refs.~[\onlinecite{AL1,AL3}], the authors of ref.~[\onlinecite{TSA}] gave a qualitative explanation of the difference between the semiclassical and the random matrix methods. They argued that the crossover happens at energy of the order of the inverse Ehrenfest time, $1/\tau_{\rm E}$. Papers [\onlinecite{AB,Bnew}] presented quantitative results for the energy gap and we discuss these results in Sec.~\ref{sec:5}.

We show that in both disordered and chaotic Andreev billiards the density of states has a gap and a square root singularity above the gap. We study a disordered Andreev billiard, containing isotropic point scatters, characterized by arbitrary mean free path $l_0$. We investigate the crossover from the semiclassical\cite{LN,ScB} to quantum regime\cite{MBFB} as parameter $l_0/l_{\rm esc}$ decreases (here $l_{\rm esc}=\esc v_{\rm F}$ is the average length of trajectories in the dot and $v_{\rm F}$ is the Fermi velocity). We further demonstrate  that the random matrix result of ref.~[\onlinecite{MBFB}] for reflectionless contact is applicable only at intermediate impurity concentration, when the mean free path $l_0$ is smaller than a typical length of trajectories in the dot, $v_{\rm F}\esc$, and longer than size $b$ of the SN contact, $b\ll l_0\ll v_{\rm F}\esc$. As the mean free path becomes shorter than the SN contact size, $l_0\ll b$, we find the density of states, described by a solution of the Usadel equation.\cite{fn2}  

For the clean chaotic billiards,
we apply general methods developed in refs.~[\onlinecite{AL1,AL2,AL3,AAL}] to study the density of states. We investigate the crossover from the semiclassical limit\cite{LN,ScB}, $t_{\rm E}/\esc\to\infty$, to a random matrix limit,\cite{MBFB} $t_{\rm E}/\esc\to 0$, and calculate the energy gap $E_{\rm g}$ as a function of $t_{\rm E}/\esc$.  
We find that the energy dependence of the density of states has an oscillating component if the Ehrenfest time is finite, see also ref.~[\onlinecite{AB}]. 

We demonstrate that the density of states has different shapes for systems with different values of the strength of disorder or chaos.
In all cases, the density of states has a finite energy gap. The corresponding curves of the density of states are different and cannot be transformed into each other by change of energy scales. 
Particularly, we consider the high energy asymptote of the density of states, $\e\esc\gg 1$, and show that the crossover between the semiclassical and quantum regimes can be understood as a suppression of  repetitive trajectories.

This paper is organized as follows. In the next section we present basic parameters of the model and introduce the Eilenberger equation. In section III we consider systems with quantum disorder. In section IV we study the effect of quantum chaos on the density of states. Section V contains discussion and conclusions. 

\section{Model}
\label{sec:2}
We consider a small $d$-dimensional metal grain connected to a
superconductor, see Fig.~\ref{fig:1}. The characteristic size of
the grain is $L$ and its volume is ${\cal  V}_{\rm g}\propto L^d$, with $d$ being dimension of a system (in two dimensional
system ${\cal  V}_{\rm g}\propto L^2$ stands for the dot's area). We assume that
the size $b$ of the contact between the metal grain and the
superconductor is small, so that $L\gg b$, and denote the $(d-1)$ dimension area of the contact by $S_{\rm c}\propto b^{d-1}$. We also include a random elastic potential into our model, which determines the effect of impurities or
irregularities inside the grain and at its boundaries. The length
scale between consequent scatterings off this potential is referred
to as the mean free path $l_0=v_{\rm F}\tau_0$ and $\tau_0$ is the
mean free time. According to the relation between mean free
path $l_0$ and the size of the system $L$ we can distinguish
ballistic, $l_0\gg L$, or diffusive, $l_0\ll L$, limits.

An important energy parameter for a metal grain is the Thouless
energy, which is inversely proportional to the ergodic time
$\tau_{\rm erg}$.
The ergodic time is the characteristic time scale which determines how
long it takes for an electron to reach boundaries of the system
from arbitrary point in the grain, {\it i.e.}  to explore the whole phase space of the system. The ergodic time is different in ballistic and diffusive
systems. In ballistic systems, the
ergodic time is determined by time between collisions with the
system boundaries, $\tau_{\rm erg}\sim L/v_{\rm F}$, on the other hand
in the diffusive regime ergodic time is defined by the diffusion
time between boundaries, $\tau_{\rm erg}\sim L^2/D$,
where $D=v_{\rm F}l_0/d$ is the diffusion constant.

Another characteristic time
of the system is the escape time $\tau_{\rm esc}$,
which determines how long electrons
spend in the grain before colliding with the contact. This time
can be estimated as the ratio of the total boundary area and area of
the contact, multiplied by the ergodic time. Indeed, the ergodic
time determines time between collisions with boundaries, and the
areas ratio determines the average number of collisions with closed
boundaries before a collision with the contact takes place. We have
$\tau_{\rm esc}\propto \tau_{\rm erg}(L/b)^{d-1}$.
We quantify the escape time from the dot for a two dimensional system by considering the volume $\Gamma$ of the phase space of the grain.
The integral over the phase space of the dot can be calculated using two
different approaches. First, directly integrating over the direction of
the momentum $\n$ and the volume of the grain, we obtain
$\Gamma=2\pi {\cal  V}_{\rm g}$. On the other hand, every point in the phase space is
uniquely determined by a trajectory, starting at the superconductor normal metal contact (SN contact). The integral over the phase space in this case can be represented as
\begin{equation}
\label{1}
\Gamma=\int_{-\pi/2}^{\pi/2}
d\theta\int_{S_{\rm c}} \cos\theta
l(\theta,b) db =2 l_{\rm esc}S_{\rm c},
\end{equation}
and we obtain the escape time $\esc= l_{\rm esc}/v_{\rm F}$:
\begin{equation}
\label{2}
\gamma_{\rm esc}=\frac{1}{\esc}=\frac{v_{\rm F} S_{\rm c}}{\pi {\cal  V}_{\rm g}},
\end{equation}
where ${\cal  V}_{\rm g}$ is the area of two dimensional metal.
Below we will usually use the escape rate $\gamma_{\rm esc}=1/\tau_{\rm esc}$.

We emphasize, that the escape time $\esc$, Eq.~(\ref{2}), based on the trajectory calculation
coincides with the escape time, defined in the random matrix theory as
$\hbar/\esc=N\mls/2\pi$, where $N$ is the number of open channels in the
contact, $N=2b/\lambda_{\rm F}$, $\mls=2\pi\hbar^2/m_{\rm e}S_{\rm g}$ is the mean level spacing, $\lambda_{\rm F}$ is the Fermi wave length, $m_{\rm e}$ is the electron effective mass and $S_{\rm g}$ is the area of the dot.
In this paper we assume $\mls$ to be negligibly small and $N$ to be large.

The superconductivity order parameter $\Delta$ is constant inside the superconducting lead and vanishes fast at the contact with the metal grain. 
We consider the limit of strong superconductivity,
$\Delta\gg \gamma_{\rm esc}$, and study the density of states at energy $\e \sim \gesc$ much smaller than the energy gap $\Delta$ in the superconductor, $\e\ll \Delta$.

The electron Hamiltonian is the sum of kinetic
and potential terms:
\begin{equation}
\label{3}
H=\frac{\p^2}{2m}-\e_{\rm F}+V_{\rm s}(\rbf)+V_{\rm q}(\rbf),
\end{equation}
where the potential term $V_{\rm s}(\rbf)+V_{\rm q}(\rbf)$ is divided
into semiclassical smooth $V_{\rm s}$  part, which describes the
effect of the grain boundaries, and the quantum part, $V_{\rm
q}$, which represents impurity scattering:
$
V_q(\rbf)=\sum_i V_i\delta (\rbf-\R_i).
$
We write the full Hamiltonian in the Gorkov-Nambu space as
\begin{equation}
\label{4}
\hat H_{\rm GN}= H \hat \tau_z + \hat \Delta(\rbf),
\end{equation}
where the term $\hat \Delta(\rbf) = \hat \tau_x\Delta \tilde \theta(\rbf)$ takes into account superconductor pairing, $\hat \tau_{x,z}$ are the Pauli matrices and $\tilde \theta(\rbf)=1$ inside the superconductor and
vanishes otherwise. This approximation for the coordinate dependence of the order parameter is justified in the limit of strong superconductivity,  
$\Delta\gg\e$.

We introduce the Green's function as an inverse function of the
Hamiltonian:
\begin{equation}
\label{5}
\left(\e-\hat H_{\rm GN} \right)
\hat G(\e,\rbf,\rbf')=\delta(\rbf-\rbf')\hat 1
\end{equation}
and  perform the Wigner transformation
\begin{equation}
\label{6}
\hat G(\e,\R,{\bf p})=\int e^{i{\bf pr}}\hat G(\e,\R+\rbf/2,\R-\rbf/2) d\rbf.
\end{equation}
Next, we change coordinates in the phase space from $\rbf$ and $\p$ to $\xi$
and $\x$, where $\xi=H(\rbf,\p)-\e_{\rm F}$ and
$\x=(\rbf,\n)$ with $\n=\p/|\p|$.
The reduced Green's function is obtained by integration over $\xi$:
\begin{equation}
\label{7}
\g(\e,\x)=\left(
\begin{array}{cc}
  g(\e,\x) & f_+(\e,\x) \\
  f_-(\e,\x) & -g(\e,\x) \\
\end{array}
\right)
=\frac{i}{\pi\nu}\int \hat G(\e,\xi,\x) d\xi,
\end{equation}
where $\nu$ is the density of states in the grain per spin degree of freedom in the absence of superconductor.
The semiclassical Green's function allows us to calculate electron
density of states in the system as an integral over the phase space of electrons in the grain: 
\begin{equation}
\label{8}
\N(\e)=\nu{\rm Re} \int {\rm Tr}\left\{
[\hat \tau_z, \g(\e,\x)]
\right\}\frac{d\x}{\Omega_d}
\end{equation}
Here $[\cdot,\cdot ]$ stands for a commutator. An element of the phase space $d\x$ is $d\x=d\rbf d\Omega_{\n}$;  $d\Omega_{\n}$ denotes a measure on a sphere of a unit radius, representing electron momentum direction $\n$,  and $\Omega_d$ is the total area of a sphere of a unit radius in $d-$dimensional space, $S_2=2\pi$ and $S_3=4\pi$.

The reduced Green's function $\g(\e,\x)$ satisfies the Eilenberger equation: \cite{Eilen,LO}
\begin{eqnarray}
\left[ i\e\hat\tau_z+  \hat\Delta (\rbf), \g(\e,\x)\right]
- {\cal L}\g(\e,\x)=I[\g(\e,\x)],
\label{9}
\end{eqnarray}
where
\begin{equation}
\label{10}
{\cal L}=
\frac{\p}{m}
\frac{\partial}{\partial \rbf}-
\frac{\partial V_{\rm s}(\rbf)}{\partial \rbf}\frac{\partial}{\partial \p}
\end{equation}
is the Liouville operator and $I[\g(\e,\x)]$ is the scattering
term. The scattering term takes into
account quantum part of the potential $V_{\rm q}$, introduced
in Eq.~(\ref{3}). A general form for the scattering
term in the Bohr approximation is:
\begin{subequations}
\label{11}
\begin{equation}
\label{11a}
I[\g(\rbf,\n)]=[\hat \Sigma(\rbf,\n),\g(\rbf,\n)],
\end{equation}
where $\hat \Sigma(\rbf)$ stands for the integral
\begin{eqnarray}
\hat \Sigma(\e,\rbf,\n) & = &
\left(
\begin{array}{cc}
  \sigma_0(\e,\rbf,\n) & \sigma_+(\e,\rbf,\n) \\
  \sigma_-(\e,\rbf,\n) & -\sigma_0(\e,\rbf,\n) \\
\end{array}
\right)
\label{11b}
\\
\nonumber
& = &
\pi\nu n_{\rm i}
\int \g(\rbf,\n') |V_{\rm q}(\n'-\n)|^2 \frac{d\Omega_{\n'}}{\Omega_d}
\end{eqnarray}
\end{subequations}
over direction of momentum $\n$ at the Fermi surface, 
$V_{\rm i}(\n-\n')$ represents the scattering
amplitude off a single impurity, the impurity concentration is
$n_{\rm i}$.

A solution of the Eilenberger equation, Eq.~(\ref{9}), is
restricted by the constraints:
\begin{equation}
\label{12}
\g^2(\e,\x)=\hat 1, \ \ \ {\rm tr}\ \g(\e,\x)=0,
\end{equation}
which represent the normalization conditions for $\g(\e,\x)$.

Inside the superconductor at distance $\xi$ away from the SN contact the Green function is given by 
\begin{equation}
\g(\e,\xi)=\hat\tau_x+
\alpha_+e^{2\Delta\xi}
\left(
\begin{array}{rr}
  1 & - 1 \\
   1 & -1
\end{array}
\right)
+
\alpha_-e^{-2\Delta\xi}
\left(
\begin{array}{rr}
  1 & 1 \\
   -1 & -1
\end{array}
\right),
\end{equation} 
where $\xi<0$ for electrons moving in the superconductor towards the grain and $\xi>0$ for electrons moving in the superconductor away from the grain.
The requirement that the Green function inside the superconductor relaxes to $\g(\e,\xi)=\hat\tau_x$, gives the following form of the Green function $\g_{\rm s}(\e,\x)$
at the SN contact inside the superconductor:\cite{LN}
\begin{equation}
\label{31}
\g_{\rm s}(\e,\x)= \hat \tau_{x}  +
\alpha
\left(
\begin{array}{cc}
  1 & - \varsigma \\
   \varsigma & -1
\end{array}
\right),
\end{equation}
where $\varsigma=+1$ for electrons incoming to the grain, \emph{i.e.} $\n\n_{\rm s}>0$, and $\varsigma=-1$ for outgoing electrons, $\n\n_{\rm s}<0$. Here $\n_{\rm s}$ is a unit vector perpendicular to the superconductor-metal interface and directed to the grain. The second term of Eq.~(\ref{31}) has a specific form, so that it corresponds to a vanishing solution of the Eilenberger equation inside the superconductor.\cite{LN}
Coefficient $\alpha$ is a free parameter. For a reflectionless SN contact the Green function is continuous and Eq.~(\ref{31}) represents the boundary conditions for electron Green functions inside the grain at the SN contact.

We represent a point of the phase space by its coordinate $\rbf$ and the direction of the electron momentum $\n$. Alternatively, we can parameterize the phase space by  trajectories, which begin and end at the SN interface. Since only one trajectory goes through  point $\rbf$ in direction $\n$, the trajectory and the distance $\zeta$  along this trajectory, measured from the starting point, uniquely represent point $\x=(\rbf,\n)$ of the phase space. Below we often use the trajectory parameterization of the phase. A point at the SN contact with $\n\n_s>0$ has always $\zeta=0$, and a point at the contact with $\n\n_s<0$ has $\zeta=l$, where $l$ is the length of the corresponding trajectory. We omit the Fermi velocity and imply that the distance $\zeta$ is measured in time units. In this case, $\zeta$ corresponds to time, which electron travels from the beginning of a trajectory to a given point, and $l$ is the total time of motion along this trajectory.

The normalization conditions, Eqs.~(\ref{12}), for the semiclassical
Green's function $\g(\e,\x)$ leaves only two independent
functions, which parameterize the matrix $\g(\e,\x)$. For certain
purposes it is convenient to choose the Riccati parametrization in
terms of functions $a_{\pm}=a_{\pm}(\e,\x)$:\cite{Maki}
\begin{equation}
\label{13}
\g = \frac{1}{1+\apq\amq}
\left(
\begin{array}{cc}
1-\apq  \amq  &   2\apq  \\
2 \amq  & \apq  \amq -1
\end{array}
\right)
\end{equation}

The Eilenberger equations for functions $a_{\pm}(\e,\x)$  acquire the following form:
\begin{subequations}
\label{14}
\begin{eqnarray}
\left(2i\e-{\cal L} \right) \apq  & = & \sigma_- \apq^2 
+2 \sigma_0 \apq -\sigma_+,
\label{14a}
\\
\left(2i\e+{\cal L} \right) \amq  & = &  \sigma_+ \amq^2  + 
2 \sigma_0\amq -\sigma_-.
\label{14b}
\end{eqnarray}
\end{subequations}
Here the self energy elements $\sigma_\pm=\sigma_\pm(\e,\x)$ and $\sigma_0=\sigma_0(\e,\x)$ are defined by Eq.~(\ref{11b}).

The boundary conditions, Eq.~(\ref{31}), acquire a simple form in the Riccati
parameterization.
Using the continuity condition for the
semiclassical Green's function $\g(\e,\x)$ at the SN interface,
and comparing Eqs.~(\ref{13}) and (\ref{31}) we
find that $\apq(\e,l)=1$ at the end of a trajectory of length $l$, 
[$\varsigma<0$ in Eq.~(\ref{31})] and $\amq(\e,\zeta=l)$ is not restrained and reflects a freedom of factor $\alpha$ in Eq.~(\ref{31}). Similarly, $\amq(\e,0)=1$ for incoming electrons, $\varsigma>0$, while $\apq(\e,\zeta)$ is a free parameter, see Table~\ref{tab:1}.

\begin{table}
\caption{The boundary conditions for the functions $a_\pm(\e,\zeta)$.}
\begin{ruledtabular}
\begin{tabular}{|c|ccc|}
    & $a_+(\e,\zeta)$ & $a_-(\e,\zeta)$ & \\ \hline
   $\zeta=0$ & unrestrained  & 1 & \\
   $\zeta=l$ & 1  & unrestrained & \\
\end{tabular} 
\end{ruledtabular}
\label{tab:1}
\end{table}

\begin{figure}
\epsfxsize= 7.5 cm  
\centerline{\epsfbox{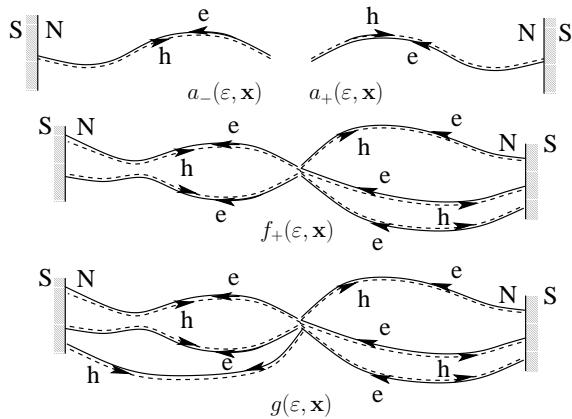}} 
\caption {
In this figure we demonstrate the meaning of the Riccati parameterization, introduced by Eq.~(\ref{13}). Functions $a_{\pm}(\e,\x)$ represent a result  of a single Andreev reflection on the SN interface. Particularly, $\amq(\e,\x)$
describes the process, in which an electron from point $\x$ travels to the SN contact and experiences an Andreev reflection, then a hole moves along exactly the same trajectory back to point $\x$. In this figure electrons, represented by a solid line, move from right to left and holes, shown by a dashed lines, move from left to right. The elements $f_\pm(\e,\x)$ and $g(\e,\x)$  may be represented in powers of the $a_{\pm}(\e,\x)$ functions. 
}
\label{fig:2}
\end{figure}

In the limit of strong superconductivity, $\Delta\gg\e$, an electron moves along some trajectory and when it is Andreev reflected as a hole, the hole moves in the opposite direction along the same trajectory.
Functions $a_{\pm}(\e,\rbf,\n)$, which describe electron Green
function in  the Riccati parametrization, Eq.~(\ref{13}), have the
meaning of the phase factor, which electron-hole pairs acquire as a result of a single Andreev reflection at the SN contact while they travel from the contact to point $\{\rbf,\n\}$ (note the boundary conditions in Table~\ref{tab:1}).
We refer to $a_{\pm}(\e,\rbf,\n)$ as wave functions of an electron-hole pairs  at point $\{\rbf,\n\}$. 
The off-diagonal elements of the Green function $\g(\e,\x)$ may be represented as a series in $a_{\pm}(\e,\rbf,\n)$ and correspond to the full amplitude, which takes into account multiple Andreev reflections, see Fig.~\ref{fig:2}. The diagonal element of the
Green function may also be written in the form of a series, which gives  us the following expression for the electron density of states in the grain:
\begin{eqnarray}
\!\! \!\! \N(\e) &\!\!  = \!\!  & \frac{2}{\mls}\int 
\frac{d\x}{{\cal  V}_{\rm g} \Omega_d}
\nonumber 
\\
& \!\! \times \!\! & \!\!
\left( 1+ 2 \sum_{k=1}^{\infty}(-1)^k
{\rm Re}\left\{\apq^k(\e,\x)\amq^k(\e,\x)\right\}
\right)
.
\label{8_s}
\end{eqnarray}
Here $\mls =\nu {\cal  V}_{\rm g}$ is the mean level spacing of electron states in the grain of the volume ${\cal  V}_{\rm g}$.
The $k$th term of this series represents the effect
of $k$ Andreev reflections. We notice, that the first term of this
expansion (unity) correspond to electron Green function in the grain without superconductivity at the contact.

In conclusion of this Section, we make the following observation. The scattering term vanishes after integration over the direction of momentum at each point of coordinate space. 
If we integrate the Eilenberger equation in the form of Eq.~(\ref{9})
over the phase space, we obtain
\begin{equation}
\label{33}
\int_{S_{\rm c}} \n \n_s \ \ \g(\e,\rbf,\n)  
dS_{\rm c} \frac{d\Omega_{\n}}{\Omega_d}= 
i\e \int_{{\cal  V}_{\rm g}} [\hat \tau_z,\g(\e,\rbf,\n)] 
d\rbf \frac{d\Omega_{\n}}{\Omega_d},
\end{equation}
where $dS_{\rm c}$ is an element of the SN interface and $\n_{\rm s}$
is a unit vector, perpendicular to $dS_{\rm c}$.
We notice that the diagonal part of Eq.~(\ref{33}) gives
\begin{subequations}
\begin{equation}
\int_{S_{\rm c}} \n \n_s \ \ g(\e,\rbf,\n)  dS_{\rm c} d\Omega_{\n} =0.
\end{equation}
The latter equation represents conservation of the electric charge in the grain.
The left hand side part of the off-diagonal part of Eq.~(\ref{33}) contains the following difference $f_+(\e,l)-f_+(\e,0)$ of the off-diagonal elements of the Green function at the terminals of trajectory of length $l$.
From the boundary conditions, defined by Eq.~(\ref{31}), we know that $f_+(\e,\zeta)$ and $g(\e,\zeta)$ are related at the trajectory terminals:
\begin{eqnarray}
f_+(\e,0)& = & 1-g(\e,0), \nonumber \\
f_+(\e,l)& = & 1+g(\e,l). \nonumber 
\end{eqnarray}
These relations allow us to write the following equation for the off-diagonal part of Eq.~(\ref{33}):
\begin{equation}
\label{34}
\int_{S_{\rm c}} |\n \n_s| \ \ g(\e,\rbf,\n) 
dS_{\rm c} \frac{d\Omega_{\n}}{\Omega_d}  =
2i \e\int_{{\cal  V}_{\rm g}} f(\e,\rbf,\n)  d\rbf \frac{d\Omega_{\n}}{\Omega_d}.
\end{equation}
\end{subequations}
This equation makes a connection between the boundary values of the
Green's function and the integral of the Green's function over the
phase space in the metal grain.

\section{Quantum Disorder}
\label{sec:3}

In this section we study density of states in the dirty system,
when the quantum part of the potential is produced by
small (point) impurities. The collision term can be written in the
form of Eqs.~(\ref{11}).

The mean free time is defined as the integral over different
directions of momentum:
\begin{equation}
\label{18}
\frac{1}{\tau_0}=2\pi\nu n_{\rm i}
\int |V_{\rm q}(\n-\n')|^2 \frac{d\Omega_{\n}}{\Omega_d}.
\end{equation}
Note, that the angular dependence of the Green's function
introduces scattering time, different from the mean free path $\tau_0$.
Particularly, in the Usadel equation angular dependence of the Green's
function on the momentum direction leads to the appearance of
the transport mean free time $\tau_{\rm tr}$,
see Eq.~(\ref{20}) below.

In this section we assume that the scattering potential has no sharp
angle dependence, and we use the mean free time as a
characteristic value of the impurity scattering strength.

\subsection{Diffusive contact}
\label{sec:3.1}

In the limit when the scattering is strong, Eilenberger equation
reduces to the Usadel equation\cite{U} [we will discuss conditions
for applicability of the Usadel equation below]. In this limit, the
Eilenberger Green's function weakly depends on the direction of
the momentum $\n$ and we represent Green's function in the form:
$\g(\e,\rbf,\n)=\g_0(\e,\rbf)+ \n \hat {\bf g}_1(\e,\rbf)$, where we assume
$\g_0(\e,\rbf)\gg \n {\bf g}_1(\e,\rbf)$. Taking into account the normalization
condition, Eq.~(\ref{12}), we find
\begin{equation}
\label{19}
{\bf g}_1(\e,\rbf)=-v_{\rm F}\tau_{\rm tr} \g(\e,\rbf)
\frac{\partial}{\partial \rbf} \g(\e,\rbf),
\end{equation}
where we introduced the transport mean free time $\tau_{\rm tr}$:
\begin{equation}
\label{20}
\frac{1}{\tau_{\rm tr}}=2 \nu n_{\rm i} \int (1-\n\n') |V_{\rm q}(\n'-\n)|^2
\frac{d^d\n}{\Omega_d}.
\end{equation}
Substituting ${\bf g}_1(\e,\rbf)$ into the Eilenberger equation,
Eq.~(\ref{9}), we obtain the Usadel equation for a system with time-reversal symmetry:
\begin{equation}
\label{21a}
D\nabla^2\Theta=2i \e\sin\Theta -2\Delta(\rbf)\cos\Theta,
\end{equation}
where we parameterize the Green  function by $\Theta=\Theta(\e,\rbf)$
\begin{equation}
\label{21b}
\g=\left(
\begin{array}{cc}
  \cos \Theta & \sin\Theta \\
  \sin \Theta & -\cos \Theta
\end{array}
\right)
\end{equation}
and $D=v^2_{\rm F}\tau_{\rm tr}/d$ is the diffusion coefficient.

\begin{figure}
\epsfxsize= 7.5 cm  
\centerline{\epsfbox{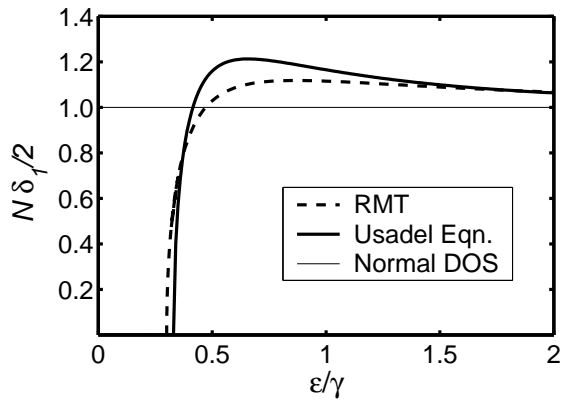}} 
\caption {
The plot shows density of states  in an Andreev billiard with strong disorder, when the mean free path $l_0$ is smaller than the size of the SN contact $b$ [solid line].
The units of the horizontal axis $\e/\gamma$ represent energy in terms of $\gamma=\gamma_{\rm c}$, see Eq.~(\ref{26}).
For comparison, we plot the density of states obtained within a random matrix theory (see ref.~\cite{MBFB} and Sec.~\ref{sec:3.2.1} of this paper)  as a function of dimensionless energy $\gamma=\gesc$, where $\gesc$ is the escape rate [dashed line].
}
\label{fig:3}
\end{figure}

We solve the Usadel equation for a  $d$-dimensional metal
grain of volume  ${\cal  V}_{\rm g}\propto L^d$, connected to a superconductor by a small contact of the area $S_{\rm c}$, see Fig.~\ref{fig:1}.
In the normal region we substitute $\Delta=0$ into Eq.~(\ref{21a}) and
integrate the resulting equation over the volume ${\cal  V}_{\rm g}$ of the grain:
\begin{equation}
\label{23}
D\int_{S_{\rm c}}\n_{\rm s} \nabla\Theta({\e,\bf s}) \ d S_{\rm c}=
2i \e\int_{{\cal  V}_{\rm g}}\sin\Theta(\e,\rbf)d\rbf,
\end{equation}
where the integral in the left hand side is taken over the area of
the contact $S_{\rm c}$, and $\n_{\rm s}$ denotes the normal vector
at the SN contact.
Since $\Theta(\rbf)$ varies only close to the contact, we neglect
variation of $\Theta$ and replace the right hand side of
Eq.~(\ref{23}) by $2\e {\cal  V}_{\rm g}\sin\Theta_0$, where $\Theta_0$ is
an asymptotic value of $\Theta(\rbf)$ away from the contact.

To calculate the left hand side of Eq. (\ref{23}) we solve the
following equation
\begin{equation}
\label{24}
D\nabla^2\Theta=0,
\end{equation}
neglecting the term with energy
$\e$ in the Usadel equation. This
approximation is justified for energy smaller than the characteristic value of the left hand side of Eq.~(\ref{24}), which we estimate as 
$D/b^2$.

We notice that Eq.~(\ref{24}) coincides with equation $\nabla^2\varphi(\rbf)=0$ for an electrostatic potential $\varphi(\rbf)$ in the problem of calculation of the current
$$
I=\int_{S_{\rm c}} \n_{\rm s} {\bf j}  d S_{\rm c}
$$
through the contact. We can write the Ohms law $I=(\varphi_{\rm c}-\varphi_0)/R_{\rm c}$ for the current $I$, where $\varphi_{\rm c}$ and 
$\varphi_0$ are the values of the electric potential $\varphi$ at the contact and inside the grain, respectively, and $R_{\rm c}$ is the resistance of the contact.
The contact resistance $R_{\rm c}$ depends on the geometry of the contact. 
For a circular contact with diameter $b$ we have $\sigma R_{\rm c}\propto 1/b$ for a three dimensional system and $\sigma R_{\rm c}\propto \log L/b$ 
for a two dimensional system.

Identifying the left hand side of Eq.~(\ref{23}) with the current $I$ and the Usadel parameter $\Theta(\rbf)$ with the electric potential $\varphi(\rbf)$, we reduce the solution of the Usadel equation to the electrostatic problem.
Then, using the boundary condition for $\Theta=\pi/2$ at the contact and introducing the value of the parameter $\Theta$ in the grain as $\Theta=\pi/2-\vartheta$, we obtain the equation for $\vartheta$: 
\begin{equation}
\label{26}
\vartheta=2i\frac{\e}{\gamma_{\rm c}} \cos\vartheta,\ \ \ \ \
\gamma_{\rm c}=\frac{D}{\sigma R_{\rm c} {\cal  V}_{\rm g}}.
\end{equation}
Here we introduced the quantity $\gamma_{\rm c}$, which has a meaning of the electron escape rate from the grain with a diffusive contact.  Because $\sigma\propto e^2\nu D/\hbar$, we have 
$\gamma_{\rm c}\propto (\hbar /e^2 R_{\rm c})(1/\nu {\cal  V}_{\rm g})=
N_{\rm trans} \mls$, 
where $N_{\rm trans}=\hbar /e^2 R_{\rm c}$ is the number of transparent channels in the SN contact [compare $\gamma_{\rm c}$ to $\gesc=N\mls/2\pi$ for a clean contact with $N$ channels].

The density of states in the grain is given by the real part of the Eilenberger Green function
\begin{equation}
\label{26.5}
{\cal N}(\e)=\frac{2}{\mls}{\rm Re}\sin\vartheta.
\end{equation}
and vanishes, if $\vartheta$ is purely imaginary. For small energy 
$\e<E_{\rm g}^{\rm U}$, Eq.~(\ref{26}) has onlu imaginary solutions, and the spectrum begins at threshold energy $E_{\rm g}^{\rm U}$, where $\vartheta$ acquires a real part. From Eq.~(\ref{26}) we find the gap energy $E_{\rm g}$
\begin{subequations}
\label{27}
\begin{equation}
\label{27a}
E_{\rm g}^{\rm U}\approx 0.331 \gamma_{\rm c}.
\end{equation}
Above the gap energy $E_{\rm g}^{\rm U}$ the density of states exhibits a square root singularity:
\begin{equation}
\label{27b}
{\cal N}(\e)=\frac{c_{\rm U}}{\mls} \sqrt{\frac{\e}{E_{\rm g}^{\rm U}}-1}
\end{equation}
\end{subequations}
and the numerical constant $c_{\rm U} \approx 3.62$. 
At higher energy the density of states approaches its metal
value $\nu$, according to the following expression:
\begin{equation}
\label{29}
{\cal N}(\e)  = \frac{2}{\mls}\left(1+\frac{\pi^2}{32} 
\frac{\gamma_{\rm c}^2}{\e^2}\right).
\end{equation}
We plot the density of states, described by Eq.~(\ref{26.5}) through a solution of Eq.~(\ref{26}), in Fig.~\ref{fig:3} by a solid line. For comparison, we present the random matrix result\cite{MBFB} for the case of reflectionless channels by a solid line.\cite{fn2}

This result is applicable for $\e\ll D/b^2$.
At larger energies, $\e\sim D/b^2$, Eq.~(\ref{24}) and, consequently, Eq.~(\ref{26}) for the parameter $\vartheta$ are no longer valid.

It is instructive to compare the result of Eqs.~(\ref{27}) with
the solution of Usadel equation in the one dimensional case for wires of length $L$ with both ends connected to a superconductor, see 
refs.~[\onlinecite{1dU,OSF}].
The value of the energy gap and density of states above the gap
are given by:
\begin{equation}
\label{30}
E_{\rm g}\approx 3.12 \frac{D}{L^2}, \ \ \ \ 
\N(\e)\approx \frac{2.30}{\mls}\sqrt{\frac{\e}{E_{\rm g}}-1}.
\end{equation}
We observe that $E_{\rm g}\sim D/L^2$. In this case the Laplasian term in Eq.~(\ref{23}) is
comparable with the energy term and approximation of Eq.~(\ref{24})
is not applicable. The Green's function is obtained as a result of
exact integration of the one-dimensional sine-Gordon equation.\cite{1dU,OSF}

Deriving Eq. (\ref{19}) we have to assume that
${\bf n}\hat {\bf g}_1\ll 1$, on  the other hand, using
Eq.~(\ref{19}) and taking into account that
$\nabla\g(\rbf)\sim 1/b$, we obtain
${\bf n}\hat {\bf g}_1\sim l_0/ b$.
The result of Eqs.~(\ref{27}) and (\ref{29})  is valid in the limit when
$l_0\ll b$. Beyond this limit the Usadel
equation is not applicable. 

For one dimensional systems, the density of states was studied
beyond the Usadel equation in refs.~[\onlinecite{PBB}], where an arbitrary
relation between the system size $L$ and the mean free path $l_0$
was considered. The conclusion of refs.~[\onlinecite{PBB}] is that there 
always exists a finite energy gap in the density of states.

\subsection{Ballistic contact}
\label{sec:3.2}

In this subsection we study the density of states in the
limit, when the mean free path is larger than the size of the
contact $l_0\gg b$. In the intermediate regime, $b\sim l_0$, the result
depends on specific geometry of the contact, and the solution of
the Eilenberger equation may be studied numerically in the spirit of refs.~[\onlinecite{PBB}].

%%%%%%%%%%%%%%%%%%%%%%%%%%%%%%%%%%%%%%%%%%%%%%%%%%%%%%%%%%%%%
%%%%%%%%%%%%%%%%%%%%%%%%%%%%%%%%%%%%%%%%%%%%%%%%%%%%%%%%%%%%%
\subsubsection{Random Matrix Limit}
\label{sec:3.2.1}
%%%%%%%%%%%%%%%%%%%%%%%%%%%%%%%%%%%%%%%%%%%%%%%%%%%%%%%%%%%%%%
%%%%%%%%%%%%%%%%%%%%%%%%%%%%%%%%%%%%%%%%%%%%%%%%%%%%%%%%%%%%%

Here we consider the limit of short mean free time 
$\tau_0\ll \esc$.
The Green's function coincides  with the Green function at the contact, see Eq.~(\ref{31}) only at small distances $\zeta\lesssim \tau_0$ or $l-\zeta\lesssim \tau_0$ along trajectories from the terminals of these trajectories.  After several scatterings, at distances larger than the
mean free path, the Green's function averages out and becomes
independent from momentum direction and coordinate. We refer to this coordinate independent part of the Green's function as a zero mode and define
it by
\begin{eqnarray}
\g_0 & = & \left(
\begin{array}{cc}
  g_0(\e) & f_0(\e) \\
 f_0(\e) & -g_0(\e)
\end{array}
\right)
\nonumber
\\
\label{35}
& = &
\frac{1}{1+a_0^2(\e)}
\left(
\begin{array}{cc}
  1-a_0^2(\e) & 2a_0(\e) \\
 2a_0(\e) & -1+a_0^2(\e)
\end{array}
\right).
\end{eqnarray}
Here the second equation
introduces the Riccati parameterization of the zero mode component
in terms of a single function $a_0(\e)$, see Eq.~(\ref{13}).

The Green function 
$\g^{\rm T}(\e,\zeta)$ near the boundaries ($\zeta=0,l$) is:
\begin{eqnarray}
\!\!\!\! \g(\e,0) & = & \g^{\rm T}(\e,l)
=\left(
\begin{array}{cc}
 g(\e,0) & f_+(\e,0) \\
 f_-(\e,0) & -g(\e,0)
\end{array}
\right)
\nonumber
\\
 & =\!\!&\!\!
\frac{1}{1+a_0(\e)}
\left(
\begin{array}{cc}
  1-a_0(\e) & 2a_0(\e) \\
 2 & -1+a_0(\e)
\end{array}
\right)\! .
\label{37}
\end{eqnarray}
Here we took into account the boundary conditions 
$\amq(\e)=1$ for incoming electrons and $\apq(\e)=1$ for outgoing electrons, see Table~\ref{tab:1}. The unrestrained values of $\apq(\e)$ and $\amq(\e)$ at the SN contact coincide with the zero mode value $a_0(\e)$ inside the grain.

At sufficiently small values of the mean free time $\tau_0$, we
neglect the contribution to the average value of the Green's function from the parts of trajectories close to the SN contact, where $\g(\e,\zeta)$ is given by Eq.~(\ref{37}), since this contribution is as small as 
$ \gesc \tau_0$, where $\gamma_{\rm esc}$ is defined by Eq.~(\ref{2}) in terms of
the contact area and the grain volume.
Then, the integral equation (\ref{34}) reduces to
\begin{equation}
\label{36}
\gamma_{\rm esc} g(\e,0)  = i \e f_0(\e).
\end{equation}

Substituting $f_0(\e)$ from Eq.~(\ref{35}) and $g(\e,0)$ from Eq.~(\ref{37}) into Eq.~(\ref{36}), we rewrite Eq.~(\ref{36}) in the form:
\begin{equation}
\label{38}
\gamma_{\rm esc}\frac{1-a_0(\e)}{1+a_0(\e)}=i\e
\frac{2a_0(\e)}{1+a_0^2(\e)}.
\end{equation}
We substitute $a_0(\e)=\exp(-2i\psi)$ and reduce Eq.~(\ref{38}) to
\begin{equation}
\tan \psi \cos 2\psi =\frac{\e}{\gesc}.
\label{38.5}
\end{equation}
The density of states 
\begin{equation}
\N(\e)=\frac{2}{\mls} {\rm Re}[g_0(\e)]=\frac{2}{\mls} {\rm Im \tan 2\psi}
\label{38.7}
\end{equation} 
is defined by Eq.~(\ref{35}) in terms of the
solution $a_0(\e)$ and is finite only if $\psi$ has an imaginary part.
The function in the left hand side of Eq.~(\ref{38.5}) has a maximum at certain value $\psi_{\rm m}=\arccos (\sqrt{1+\sqrt{5}}/2)\approx 0.45$ and we define 
\begin{subequations}
\label{eq:rmt}
\begin{eqnarray}
E_{\rm g}^{\rm rmt} & = & \tan \psi_{\rm m} \cos 2\psi_{\rm m} \gesc
\nonumber
\\
& =& \sqrt{\frac{5\sqrt{5}-11}{2}}\gamma_{\rm esc}\approx 0.30 \gesc.
\label{i2}
\end{eqnarray}
At small energy $\e<E_{\rm g}^{\rm rmt}$, Eq.~(\ref{38.5}) has real solutions and the density of states vanishes.  Above the gap we expand the left hand side of Eq.~(\ref{38.5}) in $\psi-\psi_{\rm m}$ as 
$\upsilon(\psi-\psi_{\rm m})^2=E_{\rm g}^{\rm rmt}-\e$, with $\upsilon\approx 2.17$,
and obtain that the density of states has a square root singularity above the gap
\begin{equation}
\label{i3}
{\cal N}(\e)=\frac{c_{\rm  rmt}}{\mls}\sqrt{\frac{\e}{E_{\rm g}^{\rm rmt}}-1}
\end{equation}
\end{subequations}
with $c_{\rm  rmt}=2\sqrt{2+4/\sqrt{5}}\approx 3.89$.

At energies larger than the escape rate $\gamma_{\rm esc}$,
the density of states approaches the normal metal value $2/\mls$,
according to
\begin{equation}
\label{41}
{\cal N}(\e)= \frac{2}{\mls} \left(1+\frac{\gamma^2_{\rm esc}}{2\e^2}
\right).
\end{equation}
The plot of the density of states is presented in Fig.~\ref{fig:3} by the
dashed line. We emphasize that the above result for the density of states, Eqs.~(\ref{38.5})-(\ref{41}) coincides with the result of
ref.~[\onlinecite{MBFB}] obtained within random matrix theory.\cite{fn1}

%%%%%%%%%%%%%%%%%%%%%%%%%%%%%%%%%%%%%%%%%%%%%%%%%%%%%%%%%%%%%
%%%%%%%%%%%%%%%%%%%%%%%%%%%%%%%%%%%%%%%%%%%%%%%%%%%%%%%%%%%%%
\subsubsection{Crossover regime}
\label{sec:3.2.2}
%%%%%%%%%%%%%%%%%%%%%%%%%%%%%%%%%%%%%%%%%%%%%%%%%%%%%%%%%%%%%%
%%%%%%%%%%%%%%%%%%%%%%%%%%%%%%%%%%%%%%%%%%%%%%%%%%%%%%%%%%%%%

Here we consider arbitrary mean free time $\tau_0\sim \tau_{\rm esc}$ due to scattering off isotropic impurities. In this case we have the
following form of the self energy:
\begin{eqnarray}
\label{42}
\hat \Sigma(\e) =\frac{1}{2\tau_0}\int \g(\e,\rbf,\n)
\frac{d\Omega_{\n}}{\Omega_d}
\end{eqnarray}
{\it i.e.} the self energy is proportional to the angle average of
the Green's function at point $\rbf$  inside the grain. For small
dots, when the Thouless energy $E_{\rm Th}$ is the largest energy
scale of the grain, we will make a conjecture, that the
averaging over direction of electron momentum is equivalent to
averaging over the full phase space of the system:
\begin{eqnarray}
\langle \g(\e,\rbf,\n)\rangle  & = & \frac{1}{{\cal  V}_{\rm g}}
\int \g(\e,\rbf,\n) d\rbf \frac{d\Omega_{\n}}{\Omega_d}
\nonumber
\\
& = & \int \g(\e,\rbf,\n)\frac{d\Omega_{\n}}{\Omega_d}.
\label{43}
\end{eqnarray}
Indeed, different directions of the momentum represent  contributions
from different points of the grain, since we assume that the
Green's function does not change fast at short parts of trajectories, corresponding to electron motion during the ergodic time 
$\tau_{\rm erg}=1/E_{\rm Th}$. 

\begin{figure}
\centerline{\epsfxsize=8cm
\epsfbox{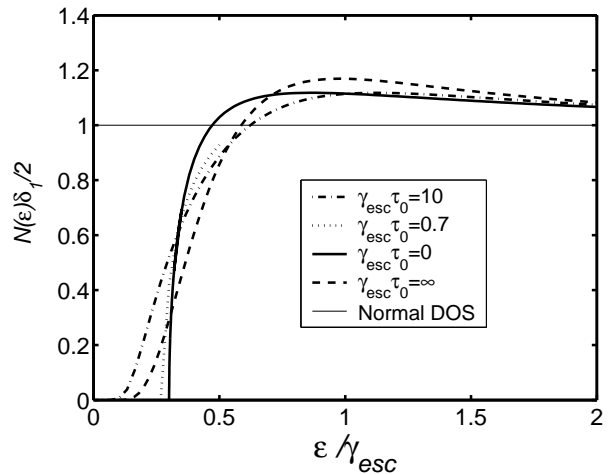}}
\caption {
The density of states for different values of mean free time $\tau_0$ is presented. In dirty metal grain ($\tau_0\gesc\ll 1$) the density of states is described by a random matrix theory [solid line]. As the strength of disorder decreases, the gap size decreases and the rise of the density of states smoothes. The cases of moderate disorder, $\tau_0\gesc=0.7$, and weak disorder, $\tau_0\gesc=10$, are shown by a dotted and dash-dotted lines respectively. [The density of states for $\tau_0\gesc=0.7$ is shown only near the gap.]  Only  in the semiclassical limit the density of states has no gap, although it is exponentially suppressed at small energy.
}
\label{fig:4}
\end{figure}

The Green's function $\g(\e,\x)$ satisfies the Eilenberger
equation in the form:
\begin{subequations}
\label{44}
\begin{equation}
\label{44a}
{\cal L}\g(\e,\rbf,\n)=[\hat H_{\rm eff}(\e),\g(\e,\rbf,\n)],
\end{equation}
where
\begin{equation}
\label{44b}
\hat H_{\rm eff}=
\left(
\begin{array}{cc}
  G_0 & F_0 \\
  F_0 & -G_0 \\
\end{array}
\right)= i\e\hat\tau_z+\frac{1}{2\tau_0}\langle \g(\e,\rbf,\n)
\rangle.
\end{equation}
\end{subequations}
Assuming that $\hat H_{\rm eff}(\e)$ is fixed, Eq.~(\ref{44a}) is linear in $\g(\e,\rbf,\n)$. We choose the
parametrization of the grain's phase space in terms of the
classical trajectories, which go through space point $\rbf$ and in
the direction of $\n$. Then, the solution of Eq.~(\ref{44a})  can
be found for each trajectory as a function of the coordinate along the
trajectory $\zeta$, since the Liouville operator has the form 
${\cal L}=\partial/\partial \zeta$. In this trajectory 
parametrization of the phase space, the smooth part of the potential $V_{\rm s}$ is taken into account as the refraction of trajectories.

The solution to Eq.~(\ref{44a}) can be written in terms of three matrices:
\begin{eqnarray}
\label{46}
\hat \varrho_z  =  \frac{1}{D_0}\hat H_{\rm eff}; \ \ 
\hat \varrho_\pm  =  \frac{1}{2D_0}
\left(
\begin{array}{cc}
  -F_0 & G_0\pm D_0 \\
  G_0\mp D_0 & F_0 \\
\end{array}
\right).
\end{eqnarray}
Matrices $\hat \varrho_z$ and $\hat \varrho_\pm$ are related to the Pauli matrices  $\hat \tau_z$ and  $\hat \tau_\pm=(\hat \tau_x\pm i\hat \tau_y)/2$ through a rotation in the Gorkov-Nambu space, which diagonalizes the Hamiltonian $\hat H_{\rm eff}$, see Eq.~(\ref{44b}). If $\hat H_{\rm eff}$ is diagonal, matrices  $\hat \varrho_i$ coincide with matrices $\hat \tau_i$. 
A general solution of Eq.~(\ref{44a}) is
\begin{equation}
\label{47}
\g(\e,\zeta)=\alpha_z \hat \varrho_z+
\alpha_+ \hat \varrho_+e^{2D_0 (\zeta-l/2)}+
\alpha_- \hat \varrho_-e^{-2D_0 (\zeta-l/2)},
\end{equation}
where $D_0^2=G_0^2+F_0^2$ and coordinate $\zeta$ is a length along
a trajectory of length $l$. Coefficients $\alpha_z$
and $\alpha_{\pm}$ are uniquely determined by the boundary conditions,
Eq.~(\ref{31}).
We have
\begin{subequations}
\label{48}
\begin{eqnarray}
\label{48a}
\alpha_z & = & \frac{F_0\cosh D_0 l+ D_0\sinh D_0 l}
{D_0\cosh D_0 l+ F_0\sinh D_0 l},
\\
\label{48b}
\alpha_\pm & = & \frac{G_0}
{D_0\cosh D_0 l+ F_0\sinh D_0 l}.
\end{eqnarray}
\end{subequations}

The average of the Green's function is defined as the integral over the phase space, which in the trajectory parameterization of the phase space is
\begin{equation}
\label{50}
\langle \g(\e,\rbf,\n)\rangle =
\gesc \int_0^\infty dl P(l) 
\int_{0}^{l} \g_l(\e,x) d\zeta,
\end{equation}
where we first integrate $\g(\e,x)$ along a trajectory of length $l$,
and then average the result over trajectories of different length $l$
with the weight 
\begin{equation}
\label{50.5}
P(l)=\gesc \exp(-\gesc l).
\end{equation}
Here $\gesc$ is escape rate, or inverse average escape time $\esc$, defined by Eq.~(\ref{2}).

Performing the integrations, we obtain:
\begin{equation}
\label{51}
\langle \g(\e,\rbf,\n)\rangle =\frac{1}{\cosh \theta}
\left(
\begin{array}{cc}
  \eta_1 \sinh \theta - \eta_2 &
  \eta_1  + \eta_2 \sinh \theta \\
  \eta_1  + \eta_2 \sinh \theta &
  \eta_2  - \eta_1 \sinh \theta \\
\end{array}
\right),
\end{equation}
and coefficients $\eta_{1,2}$ are
\begin{subequations}
\label{52}
\begin{eqnarray}
\eta_1 & = & 1+2\sum_{k=1}^{\infty}
\frac{(-1)^k \gamma_{\rm esc}^2}{(2F_0k\cosh\theta+\gamma_{\rm esc})^2}
\tanh^{2k}\frac{\theta}{2},
\label{52a}
\nonumber
\\
\eta_2 & = &
\frac{\gamma_{\rm esc}}{F_0\cosh\theta }\tanh\frac{\theta}{2}
\nonumber
\\
& + & \frac{2\gesc}{F_0\sinh\theta} \sum_{k=1}^{\infty}
\frac{(-1)^k\gamma_{\rm esc}}{2F_0k\cosh\theta +\gamma_{\rm esc}}
\tanh^{2k}\frac{\theta}{2}.
\label{52b}
\nonumber
\end{eqnarray}
\end{subequations}
Here we introduced a shorthand notation $\theta$, defined
through $G_0=F_0\sinh\theta$ and $D_0=F_0\cosh\theta$.

In these calculations, Eqs.~(\ref{44b}) and (\ref{51}) 
represent a self consistent system of equation for $F_0$ and $G_0$ (or $\theta$). 
The solution of
this self consistency equation gives the electron density
of states in the grain:
\begin{equation}
\label{54}
\N(\e)= 
\frac{2}{\mls} {\rm Re}\left\{
\eta_1 \tanh \theta - \eta_2 \frac{1}{\cosh \theta}
\right\}.
\end{equation}

\begin{figure}
\epsfxsize= 7.5 cm  
\centerline{\epsfbox{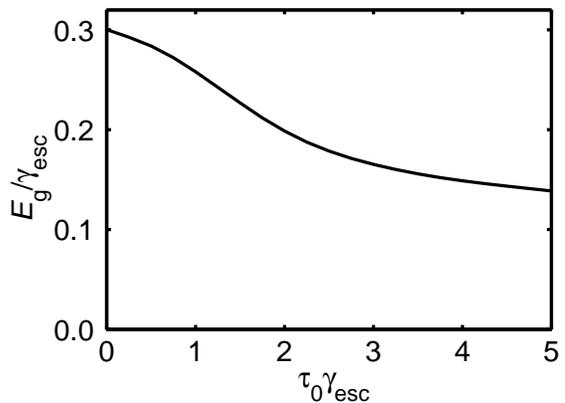}} 
\caption {
The gap energy $E_{\rm g}$ as a function of the mean free time $\tau_0$ is shown by a solid line. 
}
\label{fig:5}
\end{figure}

The density of states, in the form of Eq.~(\ref{54}), supplied with Eqs.~(\ref{44}) and (\ref{51}), are obtained here for the first time. 
These equations can be solved numerically for an arbitrary value of $\tau_0\gesc$. In Fig.~\ref{fig:4} we plot $\N(\e)$ for weak disorder, $\tau_0\gesc=10$, and for moderately strong disorder, $\tau_0\gesc=0.7$, by dash-dotted and dotted lines, respectively.
For weak disorder, the density of states is similar to the semiclassical result [dashed line], although it exhibits the gap 
$E_{\rm g}\approx 0.09 \gesc$. The gap is not noticeable in Fig.~\ref{fig:4}, since the overall number of states near the gap is already exponentially small. As the strength of disorder increases, $\tau_0^{-1}\gtrsim \gesc$, the density of states acquire the form, similar to the random matrix result [solid line], see the dotted line in Fig.~\ref{fig:4}  for the $\tau_0\gesc=0.7$ case shown only for $\e< 0.5\gesc$. 
We also present the energy gap $E_{\rm g}$ as a function of the mean free time $\tau_0$. As $\tau_0$ increases, the gap size monotonically decreases, see Fig.~\ref{fig:5}.

Now we show, that both the semiclassical result\cite{LN,ScB} and the random matrix result\cite{MBFB} may be obtained from Eqs.~(\ref{44}), (\ref{51}) and (\ref{54}).

In the absence of impurity scattering, $\tau_0\to \infty $, we obtain the
ballistic semiclassical result. Indeed, we have $G_0=D_0=i\e$,
$F_0\to 0$ and $\theta\to \infty$. The density of states is determined
by $\eta_1$:
\begin{equation}
\label{55}
\N(\e)=\frac{2}{\mls} \sum_{k=-\infty}^{\infty}
\frac{(-1)^k \gamma^2_{\rm esc}}{(2ik\e+\gamma_{\rm esc})^2}.
\end{equation}
This result is equivalent to the semiclassical expression 
for the density of states in the Andreev grain:
\begin{equation}
\label{i1}
{\cal N}(\e)=\frac{2}{\mls}\frac{\pi^2\gamma_{\rm esc}^2}{4\e^2}
\frac{\cosh \pi\gamma_{\rm esc}/2\e}
{\sinh^2 \pi\gamma_{\rm esc}/2\e},
\end{equation}
which was derived in ref.~[\onlinecite{ScB}].
At large energy, the density of states in the clean semiclassical
system approaches its value in the normal state according to
\begin{equation}
\label{56}
\N(\e)=\frac{2}{\mls}\left(1+\frac{\pi^2}{24}
\frac{\gamma^2_{\rm esc}}{\e^2}\right).
\end{equation}

In the opposite limit, $\e\tau_0\ll 1$ and $\gamma_{\rm
esc}\tau_0\ll 1$, we neglect series in Eq.~(\ref{44b})  for
$\eta_{1,2}$, since the series are of the second order in
$\gamma_{\rm esc}\tau_0$. To the zeroth order in $\e\tau_0$ and
$\gamma_{\rm esc}\tau_0$ the self-consistency equation,
Eq.~(\ref{44b}), is satisfied by arbitrary $\theta$. 
Keeping the first order terms in $\e\tau_0$
and  $\gamma_{\rm esc}\tau_0$, we obtain the equation for
$\theta$:
\begin{equation}
\label{57}
i\e =  \gamma_{\rm esc} \cosh\theta\ \tanh\frac{\theta}{2} 
\end{equation}
and the expression for the density of states
\begin{equation}
\label{57den}
\N(\e) =  \frac{2}{\mls}{\rm Re} \left\{\tanh\theta \right\}. 
\end{equation}
Equations (\ref{57}) and (\ref{57den}) are equivalent to Eqs.~(\ref{38.5})
and (\ref{38.7}) after the substitution $\theta=2i\psi$.
We conclude, that the density of states to the lowest order in
$\e\tau_0$ and   $\gamma_{\rm esc}\tau_0$ corresponds to 
the random matrix theory result, see Eqs.~(\ref{eq:rmt}) and (\ref{41}) 
and ref.~[\onlinecite{MBFB}].

The most drastic difference between the density of states corresponding to different values of the mean free time $\tau_0$ appears at small energy, see Figs.~\ref{fig:4} and \ref{fig:5}.  This difference was discussed in refs.~[\onlinecite{ScB,LN,MBFB,TSA,AB,Bnew}]. 
Here we emphasize that the difference exists even at  high
energy. Comparing Eqs.~(\ref{41}) and (\ref{56}), we notice
that already the lowest order in  $\gamma_{\rm esc}/\e$ term differs 
by a numerical factor of $\pi^2/12$. In general, a high energy asymptote 
$\e \gg \gesc$ of the density of states can be written in the form:
\begin{subequations}
\label{58}
\begin{equation}
\label{58a}
\N(\e)\approx \frac{2}{\mls}
\left(
1+\varphi(\e\tau_0) \frac{\gamma^2_{\rm esc}}{2\e^2}
\right),
\end{equation}
where $\varphi(\e\tau_0)$ is
\begin{equation}
\label{58b}
\varphi(y)={\rm Re}\left\{
\frac{1+2i-2y-\pi^2 y^2/12}{(1+i y)^2}
\right\}.
\end{equation}
\end{subequations}
We observe, that transition from the dirty limit, $\e\tau_0\ll 1$, to the clean limit, $\e\tau_0\gg 1$,
occurs at energy of the order of the scattering rate, 
$\e\sim 1/\tau_0$. 

We discuss the difference between the high energy limits of the semiclassical and the random matrix theory\cite{fn1} results in more details. Mathematically, the different coefficient in front of $\e^2/\gesc^2$ in Eqs.~(\ref{41}) and (\ref{56}) may be attributed to $\sum (-1)^k/k^2$ over $k$, when the random matrix theory result is obtained only from the first term of the sum, while the semiclassical result corresponds to 
$\sum_{k=1}^{\infty} (-1)^k/k^2=-\pi^2/12$. We argue that this mathematical difference between Eqs.~(\ref{41}) and (\ref{56}) has a physical meaning.

In the semiclassical limit, the energy levels in the grain are usually  interpreted as standing waves along classical trajectories.\cite{LN,ScB} Alternatively the contribution to the density of states may be explained as the effect of interference of repetitive motion of electron-hole pairs along a classical trajectory, see Eq.~(\ref{55}). A scattering off impurities switches electrons and holes between different trajectories and destroys the interference effect of such repetitive motion. Due to the scattering, an electron-hole pair travels along any trajectory only once, and the sum over $k$ in Eq.~(\ref{55}) -- the number of repetitions  -- is limited to $k=\pm 1$.

\section{Quantum Chaos}
\label{sec:4}

In this Section we consider a clean metal grain with chaotic smooth potential and consider the effect of quantum diffraction on electron density of states.

\subsection{Qualitative discussion}
\label{sec:4.1}

\begin{figure}
\epsfxsize= 7.5 cm  
\centerline{\epsfbox{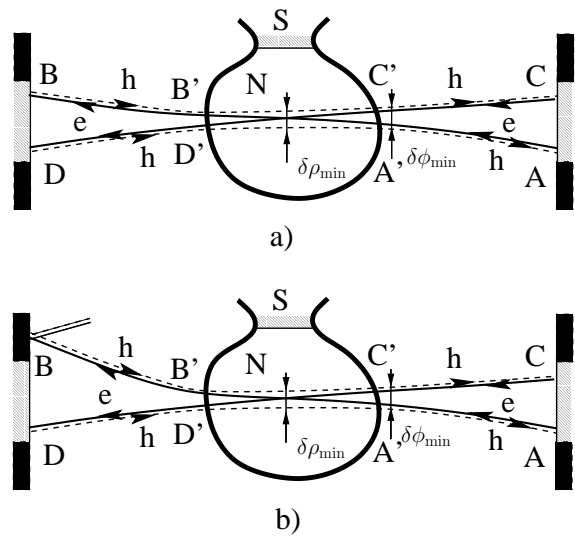}} 
\caption {
Figure a) represents $\apq^2(\e,\x)\amq^2(\e,\x)$ for the case, when the interference exists. The low figure corresponds to the in dependent propagation of two electron-hole pairs. Indeed, if one pair is Andreev reflected at point $D$, the other is normal reflected at point $B$ and the paths, along which these two pairs travel from points $B'$ and $D'$, are different by a length of order of $l_{\rm esc}$. The paths $AA'$, $BB'$, $CC'$ and $DD'$ are straitened and the electron-hole pairs have multiple reflections on the dot normal boundaries while they are moving along these paths.
}
\label{fig:6}
\end{figure}

In quantum mechanics electron propagation is described by the evolution of a wave packet. The electron Green function, see Eq.~(\ref{7}), at some point $\x$ of the phase space is actually determined by averaging over a set of wave functions, which constitute a wave packet with center of mass at point $\x$. In this sense, the effect of quantum diffraction is similar to the effect of disorder. We emphasize, however, that the quantum diffraction mixes wave functions of electron-hole pairs at close points of the phase space within a minimal wave packet, see below. On the contrary, scattering off point impurities in disordered system mixes electron-hole wave functions propagating in completely different directions. Thus, the scattering off point impurities results in the mixing of distant points of the phase space.

First we estimate the size of a typical wave packet in the phase space and then discuss the effect of chaotic dynamics on the electron Green function.

We consider a wave packet, which consists of two electron-hole pairs, travelling from points $C'$ and $A'$  to points $B'$ and $D'$ in Fig.~\ref{fig:6}. The width of the wave packet is  $\delta\rho$ and pairs' momenta are misaligned within angle $\delta \phi$, so that the momentum distribution of the wave packet has width $p_\bot=p_{\rm F}\delta \phi$, where $p_{\rm F}=2\pi\hbar /\lambda_{\rm F}$ is the Fermi momentum and 
$\lambda_{\rm F}$ is the Fermi wavelength. According to the uncertainty principle, we have $p_\bot \delta\rho\sim 2\pi\hbar$. Taking into account the  relation $\delta \rho\approx L\delta\phi$, see Fig.~\ref{fig:6}, we obtain 
\begin{equation}
\phi_{\rm min}=\sqrt{\lambda_{\rm F}/L},
\ \ \ \ \
\rho_{\rm min}=\sqrt{\lambda_{\rm F}L}
\label{eq:minpack}
\end{equation}
for the momentum deviation angle   and for the coordinate width of a minimal wave packet, respectively. 

The Eilenberger equation, Eq.~(\ref{9}), for a clean system does not describe quantum diffraction. The way to mimic quantum diffraction is to introduce small angle scattering potential, see refs. [\onlinecite{AL1,AAL}].  This scattering introduces mixing, or
``interaction'',  between electron-hole wave functions, which approach each other within a small distance in the phase space, described by Eqs.~(\ref{eq:minpack}). The small angle scattering suppresses the oscillating components of the Green functions and destroys the semiclassical description of electron-hole motion in the system.

We develop the application of the method of small angle potential to calculations of density of states in Andreev billiard in Appendix A, using the technique developed in ref.~[\onlinecite{AL1}]. Below in the text we utilize somewhat more intuitive approach. We represent the Green function in the Riccati parameterization, see Eq.~(\ref{13}) and rewrite it in terms of well defined series in $a_{\pm}^k(\e,\x)$. 
To include the effect of quantum diffraction, we identify powers of $a_{\pm}(\e,\x)$ as weighted average over the minimal wave packet of the electron-hole pairs, see Fig.~\ref{fig:6}:
\begin{equation}
a_{\pm}^k(\e,\x) \to \langle a_{\pm}^k(\e,\x) \rangle=
\prod_{i=1}^{k}\int P(\x-\x_i) a_{\pm}(\e,\x_i)d\x_i.
\label{ak-average}
\end{equation}
Function $P(\x-\x')$ represents the coordinate and momentum distribution within a minimal wave packet and vanishes fast on coordinate scale of $\rho_{\rm min}$ or for momentum deviation on angle larger than $\delta\phi_{\rm min}$, see Eqs.~(\ref{eq:minpack}). A detailed form of the function $P(\x-\x')$ is not important, since, as we show below, the density of states depends logarithmically on the width of the function $P(\x-\x')$.

In chaotic system the distance between two trajectories grows
exponentially in time $t$: 
$\rho(t)=\rho_{\rm min}\exp(\lambda t)$ and 
$\phi(t)=\phi_{\rm min}\exp(\lambda t)$, where $\lambda>0$ is the Lyapunov exponent of the classical chaotic motion. Consequently, even  
though the size of the wave packet is initially small, time evolution leads to the spreading of the wave packet over the phase space. Generally, the spreading of the wave packet is characterized by the Ehrenfest time 
$t_{\rm E}$. The Ehrenfest time shows how long it takes for a wave packet to spread over the whole phase space of the system and is equal to 
\begin{equation}
t_{\rm E}=\frac{1}{\lambda} \ln \frac{L}{\lambda_{\rm F}}=
\frac{2}{\lambda} \ln \frac{L}{\rho_{\rm min}}.
\label{eq:tE-stan}
\end{equation} 
We argue, however, that the relevant time scale for the problem of proximity effect in Andreev billiards depends on the size of the contact $b$ and is given by 
\begin{equation}
\tte=\frac{1}{\lambda} \ln \frac{b}{\rho_{\rm min}}=
\frac{t_{\rm E}}{2}-\frac{1}{\lambda}\ln \frac{L}{b}.
\label{eq:tte}
\end{equation} 

The second term in the right hand side of Eq.~(\ref{eq:tte}) may be omitted, since it contains a logarithm of macroscopic parameters $b$ and $L$.
Nevertheless we explain the appearance of the scale $b$ as follows.
If electron propagates from points $B'$ or $D'$ along trajectories $B'B$ or  $D'D$, see Fig.~\ref{fig:6}a, for time smaller than $\tte$, the spreading of the initial wave packet is smaller than the size $b$ of the SN contact and the two trajectories ends nearly simultaneously at points $B$  and $D$. We conclude that the contribution of electron-hole wave functions to the electron Green function originates from coherent motion of electron-hole pairs along these two  trajectories. This coherent contribution may be  viewed as a repetitive motion along the same trajectory, see the text at the end of Sec.~\ref{sec:3.2.2}. 

On the other hand, if electron propagates along $B'B$ and $D'D$ for time longer than $\tte$, the wave packet spreads on the scale larger than the size $b$ of the SN contact. As shown in Fig.~\ref{fig:6}b, although the electron-hole pair travelling  along $D'D$ experiences Andreev reflection at point $D$, the other pair travels along $B'B$ and misses the SN contact.  It continues its motion further from point $B$ and explores the phase space, different from the one reached by the $D'D$ trajectory. Thus, even though the wave packet originates from the contribution of electron-hole wave functions at close points in the phase space, the corresponding wave functions are quite different, since they represent motion in distant parts of the phase space. The averaging over the wave packet leads to appearance of a component of the Green function, produced from different parts of the system phase space, known as a ``zero mode''.   

We expect that if the Ehrenfest time $\tte$ is much longer than the  time of electron propagation along a typical trajectory, then the electron Green function may be determined by semiclassical methods, see refs.~[\onlinecite{LN,ScB}], while in the opposite limit of short $\tte$, the quantum diffraction destroys the repetitive modes and leads to strong coupling between electron-hole wave functions from distant parts of the phase space.
In this sense, motion of electron-hole pairs in systems with short 
$\tte\ll \esc$ actually coincides with motion in disordered systems, and the random matrix result\cite{MBFB} describes the density of states, see Sec.~\ref{sec:3.2.1}.

\subsection{Quantitative analysis}
\label{sec:4.q}
In this Subsection we perform an  analytical analysis of the density of states in clean chaotic Andreev billiards. First we consider the high energy limit $\e\gg \gesc$, when the perturbation theory is applicable. In the perturbation theory we identify the relevant features of the Eilenberger equation for a description of motion of electron-hole pairs in the regime of quantum chaos. Next, we construct a self-consistency equation, which determines the density of states at arbitrary energy $\e\sim \gesc$.

\subsubsection{High energy limit}
\label{sec:4.2} 

As we showed in section \ref{sec:3.2.2}, 
the density of states of disordered systems approaches the normal density of states as $\gamma^2_{\rm esc}/\e^2$ at large energy $\e\gg\gamma_{\rm esc}$, but the coefficient of this asymptote depends on the strength of disorder, characterized by the  mean free time $\tau_0$, see Eq.~(\ref{58b}). A similar asymptotic behavior exists in quantum chaotic systems and now the coefficient depends on the value of the Ehrenfest time $\tte$.  To prove this statement, we calculate the density of states at high energy ($\e\gg\gesc$).

We use the Riccati parameterization of the Green function and apply  Eq.~(\ref{ak-average}) for calculation of powers $a_{\pm}(\e,\x)$ at coincident points, which appear in the expression for the density of states, see Eq.~(\ref{8_s}). A more formal derivation with Perron-Frobenius differential operator is presented in Appendix A.

The density of states is determined by the sum of $\overline{\langle\apq^k(\e,\x)\rangle\langle \amq^k(\e,\x)\rangle}$ over all positive integer $k$. Here $\overline{( \dots)}$ stands for the averaging over the phase space of electrons in the grain, while $\langle \dots \rangle$ denotes averaging within a minimal wave packet according to Eq.~(\ref{ak-average}).
We observe that $\apq^k(\e,\x)$ and $\amq^k(\e,\x)$ are mutually independent, since the corresponding electron-hole pairs move in completely different parts of the phase space, approaching point $\x$ from two opposite directions (notice the different signs before the Liouville operators in Eqs.~(\ref{14})).  Due to independence of 
$\apq^k(\e,\x)$ and $\amq^k(\e,\x)$, we write
\begin{equation}
\label{63}
\overline{\langle \apq^k(\e,\x)\rangle \langle \amq^k(\e,\x)\rangle }
=\langle\overline{\apq^k(\e,\x)}\rangle 
\langle \overline{\amq^k(\e,\x)}\rangle ,
\end{equation}
so that the density of states in the grain is determined by the product of $\langle \overline{\apq^k(\e,\x)}\rangle$ and 
$\langle \overline{\amq^k(\e,\x)}\rangle$.

We start with calculation of the averages of electron-hole wave functions $\overline{a_{\pm}(\e,\x)}$. We integrate both sides of Eq.~(\ref{14}) over the phase space of electrons in the grain ${\cal V}_{\rm g}\Omega_d$. As we already mentioned, the scattering term vanishes after integration, see Eq.~(\ref{33}).
The integral of ${\cal L}_{\rm cm}a_\pm(\e,\x)$
reduces to the values of $a_{\pm }(\e,\x)$ at the SN contact. At $\e\gg\gesc$, only the contribution from the trajectory terminal with restrained value of $a_\pm(\e,\zeta)$ is important, since the unrestrained value of $a_\pm(\e,\zeta)$ at the SN contact oscillates fast and thus averages out.
We obtain
\begin{equation}
\label{69}
\overline{a_{\pm}(\e,\x)}\approx \frac{\gesc}{2i\e}.
\end{equation}

Next, we calculate the contribution to the density of states from the $k=2$ term in Eq.~(\ref{8_s}), $\langle\overline{\apq^2(\e,\x)}\rangle$,  to the second order in $\gesc/\e$. We show that this term behaves differently in semiclassical and quantum mechanics. Particularly, the quantum diffraction destroys the semiclassical contribution to the density of states, originating from the repetitive motion.

A similar problem was solved in ref.~[\onlinecite{AL1}], where the weak localization correction to the conductance of quantum chaotic systems was studied. It was shown that the weak localization correction originates as a result of quantum diffraction which couples Cooperon and diffuson modes. 
Following the technique, developed in refs.~[\onlinecite{AL1,AAL}], we introduce new functions $A_2^{\pm}(\e,\x_1,\x_2)$ with arbitrary separated points $\x_1=(\rbf_1,\n_1)$ and $\x_2=(\rbf_2,\n_2)$:
\begin{equation}
A_2^{\pm}(\e,\x_1,\x_2)=a_{\pm}(\e,\x_1)a_{\pm}(\e,\x_2).
\label{A2}
\end{equation} 

We are interested in the limit of small relative angle $\phi$ between vectors $\n_1$ and $\n_2$ within a minimal wave packet, Eq.~(\ref{eq:minpack}).  
We change variables from $\x_1$ and $\x_2$ to the coordinates of the center
of mass $\R=(\rbf_1+\rbf_2)/2, \ {\bf N}\approx (\n_1+\n_2)/2$ 
(for $|\phi| \ll 1$) and to the relative coordinates ${\bm{\rho}}=\rbf_1-\rbf_2$ and $\phi$. Under the condition 
$|\phi| \ll 1$, the vector of relative distance  ${\bm{\rho}}$ is perpendicular to the direction of ${\bf N}$ and  we refer to it through its projection $\rho$ on the axis perpendicular to ${\bf N}$.
At $|\phi| \ll 1$ the equation for $A_2^{\pm}(\e,\x_1,\x_2)$ in new variables is
\begin{eqnarray}
\label{64}
\left(
4i\e\mp {\cal L}_{\rm cm}-{\cal L}(\rho,\phi)
\right)
A_2^{\pm}(\e,\R,{\bf N},\rho,\phi)=0,
\end{eqnarray}
where 
\begin{subequations}
\label{65}
\begin{equation}
\label{65a}
{\cal L}_{\rm cm} =v_{\rm F} {\bf N}\frac{\partial}{\partial \R}-
\frac{1}{p_{\rm F}}\frac{\partial U_{\rm s}}{\partial \R}
\frac{\partial }{\partial {\bf N}}
\end{equation}
is the part of the Liouville operator describing the motion of the center of mass, and 
\begin{equation}
\label{65b}
{\cal L}(\rho,\phi) =v_{\rm F} \phi \frac{\partial}{\partial \rho}-
\frac{1}{p_{\rm F}}\frac{\partial^2 U_{\rm s}}{\partial \R^2_\bot}
\rho \frac{\partial }{\partial \phi}
\end{equation}
\end{subequations}
is the Liouville operator of relative motion of two electron-hole pairs.

We average $A_2^{\pm}(\e,\R,{\bf N},\rho,\phi)$ over the position of center of mass ${\bf N}$ and $\R$ in the phase space.
Integration of the second term with the Liouville operator ${\cal L}_{\rm cm}$ by parts gives values of $A_2^{\pm}$ at the boundaries. Particularly, due to the boundary conditions, see Table~\ref{tab:1}, for outgoing electrons $a_{\pm}(\e,\zeta=l)=1$ and we have $A_{2}^{+}(\e,\R,{\bf N},\rho,\phi)=1$ at the SN contact, provided ${\bf N}\n_{\rm s}<0$ ($\n_{\rm s}$ is a normal vector to the SN inerface). In the opposite limit ${\bf N}\n_{\rm s}>0$, the value of $A_{2}^{+}(\e,\R,{\bf N},\rho,\phi)$ is unrestrained. Similar analysis provides values of $A_{2}^{-}(\e,\R,{\bf N},\rho,\phi)$ at the SN contact.
We introduce notation 
$s_2=\langle A_2^{\pm}\rangle_{\rm sn}/\overline{A_2^{\pm}}$ for the ratio of the average unrestricted  value $A_2^{\pm}$ at the SN contact, 
$\langle A_2^{\pm}\rangle_{\rm sn}$,  to the average of $A_2^{\pm}$ over the whole phase space. 
We obtain
\begin{equation}
\label{Eqs-for-A2}
\left(4i\e+\frac{s_2}{\esc} \mp {\cal L}(\rho,\phi)\right)
A_2^{\pm}(\e,\rho,\phi)
=\frac{1}{\esc}.
\end{equation}
At high energy  $s_2\approx 1$.

\begin{figure}
\centerline{\epsfxsize=7.5cm
\epsfbox{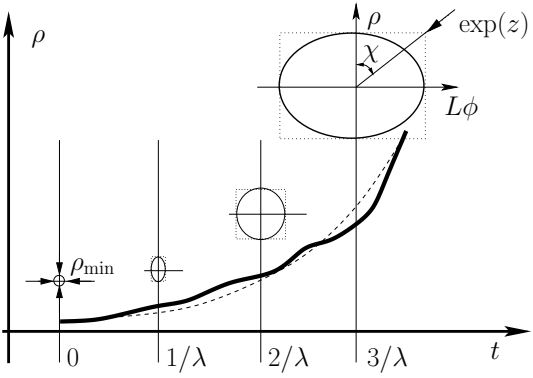}}
\caption {
Evolution of the wave packet in time.  The ellipse represents the wave packet in the phase space. The variable $z$ determines the size of the ellipse (its diagonal), while the angle variable $\chi$ determines the shape of the ellipse. The dashed line in $t$-$\rho$ plane represents the Lyapunov divergence of two trajectories $\rho(t)=\rho_{\rm min}\exp(\lambda t)$ with the Lyapunov constant $\lambda$.
}
\label{fig:7}
\end{figure}

The Liouville operator ${\cal L}(\rho,\phi)$ in Eq.~(\ref{Eqs-for-A2}) can be rewritten in new variables $z$ and $\chi$, introduced as
\begin{equation}
\label{72}
z=\frac{1}{2}\ln\frac{\phi^2L^2+\rho^2}{L^2},\ \ \ \ 
\chi=\arctan\frac{\phi L}{\rho}. 
\end{equation}
As was shown in ref.~[\onlinecite{AL1}], the variable $\chi$ is irrelevant for quantity $A_2^{\pm}(\e,\rho,\phi)$, described by a solution of Eq.~(\ref{Eqs-for-A2}).   Indeed, we represent evolution of the wave packet  in time $t$, which is a conjugate variable to energy variable $\e$ in the left hand side of Eq.~(\ref{Eqs-for-A2}), see Fig.~\ref{fig:7}. The characteristic scale of the wave packet, represented by the variable $\exp(z)$ grows in time as $\exp(\lambda t)$ with the Lyapunov exponent $\lambda$. A particular configuration of the wave packet is described by variable $\chi$, which significantly fluctuates 
as the wave packet explores the phase space and gets averaged out after integration of $A_2^\pm$ over a position of the center of mass. [Actually this averaging occurs at ergodic time scale, which is much shorter than any other time scale in the problem.] The fluctuations of the Lyapunov exponent $\lambda$ may also be taken into account. The resulting form of Eq.~(\ref{Eqs-for-A2}) is\cite{AL1}:
\begin{equation}
\left(4i\e+\gesc -\lambda\frac{\partial}{\partial z}-
\frac{\lambda_2}{2}\frac{\partial^2}{\partial z^2}\right)
A^\pm_2(\e,z)=\gesc,
\label{eq:A2-z}
\end{equation}
where $\lambda$ is the average value of the Lyapunov exponent over the electron phase space in the grain and $\lambda_2$ represents the deviation of Lyapunov exponent for different points of the phase space, see ref.~[\onlinecite{AL1}] for more details.

As the size of the wave packet reaches the size of the SN contact $b$, and $z\sim \ln b/L$, 
the $a_{\pm}(\e,\x)$ functions in $\overline{A_{2}^{\pm}(\e,\x_1,\x_2)}$ become independent. Indeed, these functions are solutions of the Eilenberger equation along different trajectories of significantly varying lengths.
We may write the following boundary condition:
\begin{equation} 
\label{68}
A_{2}^{\pm}(\e,z=\ln b/L)=\overline{a_{\pm}(\e,\x)}^2,
\end{equation} 
where $\overline{a_{\pm}(\e,\x)}$ is the average of $a_{\pm}(\e,\x)$ over the phase space and is given by Eq.~(\ref{69}). [We should distinguish the average $a_{\pm}(\e,\x)$  and the zero mode components of the Green function. More precisely, the right hand side of Eq.~(\ref{68}) contains the zero mode value of $a_{\pm}(\e,\x)$, as we will discuss in the next subsection. At high energy $\e\gg\gesc$ both quantities are equal.]

To solve Eqs.~(\ref{Eqs-for-A2}) and (\ref{68}), we represent $A_{2}^{\pm}(\e,\rho,\phi)$ as a sum of two functions: 
\begin{equation}
\label{70}
A_2^{\pm}(\e,\rho,\phi)=
A_{2,{\rm b}}^{\pm}(\e)+
A_{2,{\rm q}}^{\pm}(\e,z),
\end{equation}
where
\begin{equation}
\label{71}
A_{2,{\rm b}}^{\pm}(\e)
=\frac{\gesc}{4i\e+\gesc}\approx
\frac{\gesc}{4i\e}
\end{equation}
is a solution of Eq.~(\ref{eq:A2-z}) without any boundary conditions imposed
and $\overline{A_{2,q}^{\pm}(\e,z)}$ is a solution of homogeneous equation Eq.~(\ref{eq:A2-z}) with certain boundary conditions at  $z=\ln b/L$, so that $A_2^{\pm}(\e,z)$ in the form of  Eq.~(\ref{70}) satisfies Eq.~(\ref{68}).

A general solution of the homogeneous Eq.~(\ref{eq:A2-z}) was found in ref.~[\onlinecite{AL1}] in the limit $\gesc,\e \ll \lambda$:
\begin{eqnarray}
w(\omega,z)=\exp\left[
\left(
\frac{i\omega}{\lambda}+\frac{\lambda_2\omega^2}{2\lambda^3}
\right)z
\right],
\label{74}
\end{eqnarray}
with $\omega=4\e-i\gesc$. 
Taking into account the boundary conditions, we find 
\begin{eqnarray}
\nonumber
A_{2,{\rm q}}^{\pm}(\e,z) & = &
 \left(
\overline{a_{\pm}(\e,\x)}^2-
\overline{A_{2,b}^{\pm}(\e)}
\right)
\\
&\times &
w(4\e-i\gesc,z-\ln b/L).
\label{72.5}
\end{eqnarray}

This equation describes the Lyapunov divergence of classical trajectories, once a finite displacement between two trajectories was created by quantum diffraction. As was discussed in ref.~[\onlinecite{AL1}], Eq.~(\ref{74}) neglects the effect of quantum diffraction and a more accurate solution for the function
$w(\omega,z)$ is needed.\cite{AL1} Indeed, according to Eq.~(\ref{8_s}), the density of states contains term $A_2^{\pm}(\e,\rho,\phi)$, taken at coincident points, {\it i.e.} $\rho,\phi\to 0$. This limit corresponds to $z\to-\infty$, when $w(\e,z\to-\infty)=0$. The latter equation means that the effect of quantum diffraction, which couples Green functions at different semiclassical trajectories is not described by the Eilenberger equation without scattering term ($1/\tau_{\rm q}=0$). The scattering term modifies the equation for function $w(\e,z)$, see Appendix A for details.  We argue that instead of working in the limit $z\to -\infty$, within a logarithmic accuracy we can average quantity $A_2^{\pm}(\e,\rho,\phi)$ over $\rho$ and $\phi$ within a minimal wave packet, Eq.~(\ref{eq:minpack}). We write
\begin{eqnarray}
\Gamma_2(\e)
 & = & 
 \left< w\left(4\e-i\gesc,z-\ln \frac{b}{L}\right)\right>.
 \label{76}
\end{eqnarray}
Performing averaging over the minimal wave packet
%\begin{widetext}
\begin{eqnarray}
\left< w(\omega,z)
 %\left(\omega,z-\ln \frac{b}{L}\right)
\right>
& \sim & \int w(\omega,z) \ e^{-(\phi^2L^2+\rho^2)/2\rho^2_{\rm min}}
%\left(\omega,\ln\sqrt{\frac{\phi^2L^2+\rho^2}{b^2}}\right) 
d\phi d\rho
\nonumber
\end{eqnarray}
%\end{widetext}
with $z=\ln\sqrt{(\phi^2L^2+\rho^2)/b^2}$ we obtain
\begin{equation}
\Gamma_2(\e)=\exp\left(-\gesc\tte -4i\e\tte-\frac{8\lambda_2\e^2}{\lambda^2}\tte \right)
\end{equation}
and $\tte=(1/\lambda)\ln b/\rho_{\rm min}$ is the Ehrenfest time, defined by Eq.~(\ref{eq:tte}).
Combining Eqs.~(\ref{69}), (\ref{71}) and (\ref{72.5}) we find  
\begin{equation}
\label{77}
\langle A_{2}^{\pm}(\e,z)\rangle = \frac{\gesc}{4i\e}\left\{
1-\Gamma_2(\e)
\right\}
\end{equation}
to the lowest order in $\gesc/\e$.

Equation (\ref{77}) has a simple interpretation. Function 
$\Gamma_2(\e)$ is the
probability amplitude, that two electron-hole pairs, initially situated within the same minimal wave packet at point $\x$ of the phase space, escape the grain independently. In other words, $\Gamma_2(\e)$ is the probability amplitude that a minimal wave packet, constructed at point $\x$ expand to size $b$ of the contact before it leaves the grain. Consequently, $1-\Gamma_2(\e)$ is the
probability amplitude for the wave packet to leave the grain
before it decays due to quantum diffraction.

We generalize the above calculations of 
$\langle \overline{A_{k=2}^{\pm}} \rangle$ for arbitrary integer $k>1$. We define\cite{GM}
\begin{equation}
A_k^\pm(\e,\{\x_i\})=\prod_{i=1}^{k}a_\pm(\e,\x_i)
\label{eq:Ak-def}
\end{equation}
and write the equation for $\overline{A_k^\pm(\e,\{\x_i\})}$, averaged over the center of mass of variables $\{ x_i\}$:
\begin{equation}
\label{Eqs-for-Ak}
\left(2ki\e+\frac{1}{\esc} \mp {\cal L}(\bm{\rho},\bm{\phi})\right)
A_k^{\pm}(\e,\bm{\rho},\bm{\phi})
=\frac{1}{\esc},
\end{equation}
which is a generalization of Eq.~(\ref{Eqs-for-A2}) to arbitrary $k$. Here the variables $\bm{\rho}=\{\rho_i\}$ and $\bm{\phi}=\{\phi_i\}$ constitute a set of $2(k-1)$ variables. We argue that $2k-3$ of these variables determine the relative position of $k$ electron-hole pairs within the corresponding wave packet and thus are irrelevant,  while one variable determines the scale of the wave packet in the phase space $z=\ln(L^2\sum\phi^2_i+\rho^2_i)/2$, see Fig.~\ref{fig:7}. When the wave packet reaches the size of the SN contact, the wave functions of electron-hole pairs  $a_{\pm}(\e,\x_i)$ become independent and we have the boundary condition $z=\ln b/L$:
\begin{equation} 
\label{68k}
A_{k}^{\pm}(\e,z=\ln b/L)
=\overline{a_{\pm}(\e,\x)}^k\propto\frac{\gesc^k}{(i\e)^k}.
\end{equation} 
Following further along the steps of calculations of 
$\langle A_{2}^{\pm}(\e,z=\ln b/L)\rangle$, we obtain
to the lowest order in $\gesc/\e$
\begin{equation}
\label{77k}
\langle A_{k}^{\pm}(\e,z)\rangle =\frac{\gesc}{2ik\e}
\left\{
1-\Gamma_k(\e)
\right\}
\end{equation}
 with 
\begin{equation}
\Gamma_k(\e)=\exp\left(-\gesc\tte -2ik\e\tte-\frac{2k^2\lambda_2\e^2}{\lambda^2}\tte \right).
\label{eq:gamma-k}
\end{equation}

Using Eq.~(\ref{8_s}) we obtain the density of states at large energy, 
$\e\gg \gesc$ in the form
\begin{widetext}
\begin{equation}
{\cal N}(\e)=\frac{2}{\mls}\left[
1+\frac{\gamma_{\rm esc}^2}{2\e^2}
\left(
1- \sum_{k=2}^{\infty}
\frac{(-1)^k}{k^2}{\rm Re}
\left\{
1-\exp\left(-\left( 2 i k \e+\gamma_{\rm esc} +\frac{2k^2\lambda_2\e^2}{\lambda^2}\right)\tau_{\rm E}\right)
\right\}^2
\right)
\right].
\label{N-largeE-QC}
\end{equation}
\end{widetext}

We found that the averages of functions $\overline{\langle\apq^k(\e,\rbf,\n)\rangle}$ are
expressed in terms of $\Gamma_k(\e)$. Particularly, function $\Gamma_2(\e)$  also appears in the expressions for \emph{i)} the weak localization correction to the conductivity\cite{AL1}, \emph{ii)} level statistics correlation function\cite{AL2}, \emph{iii)} noise of a current through a chaotic cavity.\cite{AAL} 

Functions $\Gamma_k(\e)$ oscillate with a period proportional to $\tte^{-1}$.
Similar oscillations were found in the weak localization correction to the conductivity\cite{AL1} and in the energy level statistics.\cite{AL2} 
The origin of these oscillations has the following explanation.
The effect of quantum diffraction can be neglected for motion at time shorter than the Ehrenfest time. Nevertheless the diffraction effects appear nearly instantaneously, with characteristic turn-on time equal to $\lambda^{-1}$, as time of motion approaches the Ehrenfest time.
Thus we can separate two types of contributions to the Green function. One contribution represents semiclassical motion of electron-hole pairs and  is responsible for oscillations in the density of states in the phase space. 
The other contribution is a constant in phase space component of the Green function, which corresponds to the zero mode. 

\begin{figure}
\centerline{\epsfxsize=8cm
\epsfbox{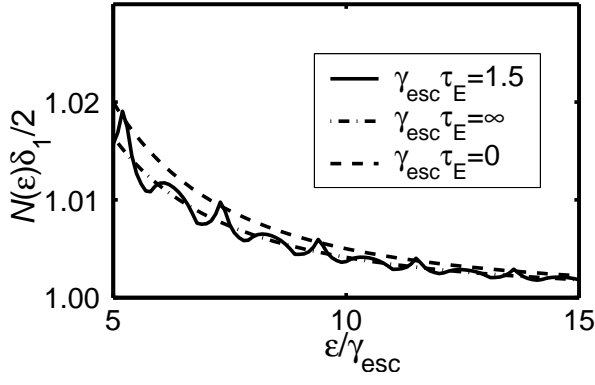}}
\caption {
The tail of the density of states at high energy, $\e\gg\gesc$.
The density of states, described by Eq.~(\ref{N-largeE-QC}) for $\tte=1.5\esc$, is shown by a solid line. The semiclassical density of states, $\tte\to\infty$ is shown by dash-dotted line, and the random matrix theory result, $\tte\to 0$, is presented by a dashed line. 
}
\label{fig:8}
\end{figure}

To illustrate the oscillations of the density of states, we plot the tail of the density of states $\N(\e)$ at large energies $\e\gg \gesc$  for  $\tte=1.5\esc$, shown by a solid line in Fig.~\ref{fig:8}.  For comparison, we also plot the density of states for the random matrix limit (dashed line) and semiclassical limit (dash-dotted line).

We emphasize that it is the sharp turn-on of the effects of quantum
diffraction at the Ehrenfest time leads to oscillations of the density of states. Fluctuations of the Lyapunov exponent suppress these oscillations. 
Indeed, the Lyapunov exponent varies for different points of the phase space and its fluctuations affect functions $\Gamma_k(\e)$ through the term
$\lambda_2\e^2\tte/\lambda^2$. 
Thus for small energies, $\e\ll \sqrt{\lambda^2/\lambda_2\tte}$, we can disregard fluctuations of the Ehrenfest time. As energy increases, the oscillations of the density of states become suppressed, see Eq.~(\ref{N-largeE-QC}). We also notice that in the quantum disorder, the process of scattering off impurities is Poissonian, and fluctuations of the free path of the order of the  mean free path itself. Consequently,
oscillations of the density of states  in disordered systems are absent, see Eqs.~(\ref{58}). 

\subsubsection{Arbitrary energy}
\label{sec:4.3}

Although the high energy expansion cannot be applied to the
calculation of the density of states at energy comparable with the
escape rate $\gamma_{\rm esc}$, exactly the same mechanism of
quantum diffraction is responsible for the appearance of the gap
in the density of states.
\begin{figure}
\epsfxsize= 7.5 cm  
\centerline{\epsfbox{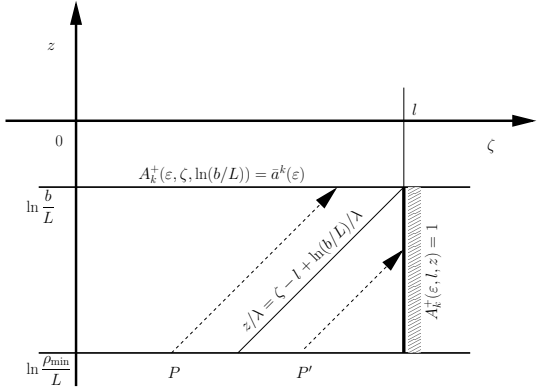}} 
\caption {
The dynamics of a minimal wave packet, initially situated at point $P$ or $P'$ is describe by the characteristics (dashed lines) of Eq.~(\ref{eq:Ak}) are presented in $\zeta-z$. The boundary conditions are defined at the terminal of the characteristics. If characteristic intersect $z=\ln b/L$ at $\zeta<l$, we say that the electron-hole wave functions at the intersection are described by a zero mode solution, $A_k^+(\e,\zeta,\ln b/L)=\bar a^k(\e)$, while if the characteristics ends at the SN contact, the boundary condition is  $A_k^+(\e,l,z)=1$.
}
\label{fig:9}
\end{figure}

Equation for $A_k^{\pm}(\rho,\phi)$ becomes significantly simplified after averaging over the phase space of the grain. Nevertheless,  this averaging happens on time scale of  the order of ergodic time $\tau_{\rm erg}$. If we are interested in the properties of the spectrum at energy much smaller than the Thouless energy $E_{\rm Th}$ (the latter is equal to the inverse ergodic time, $E_{\rm Th}=\tau^{-1}_{\rm erg}$), we can assume that the motion of electron-hole pairs constituting $A_k^{\pm}(\rho,\phi)$, is described by equation, averaged over irrelevant coordinates:
\begin{equation}
\left(
2i\e k -\frac{\partial }{\partial \zeta}-\lambda\frac{\partial }{\partial z}
-\frac{\lambda_2}{2}\frac{\partial^2 }{\partial z^2}
\right)A_k^{+}(\e,\zeta,z)=0.
\label{eq:Ak}
\end{equation}
Here $\zeta$ stands for a coordinate of the center of mass of a wave packet.
As we already discussed, the term $\partial^2/\partial z^2$ represents fluctuations of the Laypunov exponent and is important at large energy $\e\sim\sqrt{\lambda/\tte}$. Below we neglect this term. 

The function $A_k^{+}(\e,\zeta,z)$ is defined  at $0<\zeta<l$, where $l$ is the length of a trajectory, and has the boundary condition $A_k^{+}(\e,l,z)=1$ at $\zeta=l$, see Table~\ref{tab:1}. The other boundary condition exists at $z=\ln b/L$, which represents the statistical independence of electron-hole wave functions, separated by a distance $b$. We say about these functions that they correspond to a zero mode $\bar a$. Thus the boundary condition is 
$A_k^{\pm}(\e,\zeta,z=\ln b/L)=\bar a^k$, see Fig.~\ref{fig:9}. 

We write the solution to Eq.~(\ref{eq:Ak}) as
\begin{subequations}
\label{eq:sol-Ak}
\begin{eqnarray}
A_k^{+}(\e,\zeta,z) & = & 
\bar a^k e^{2ik\e(z-\ln b/L)/\lambda}
\theta\left(\frac{z-\ln b/L}{\lambda}-\zeta+l\right)
\nonumber
\\
& + &
e^{2ik\e(\zeta-l)}
\theta\left(\zeta-l-\frac{z-\ln b/L}{\lambda}\right).
\label{eq:sol-Ak1}
\end{eqnarray}
Repeating the above analysis for $A_k^{-}(\e,x,z)$, we obtain
\begin{eqnarray}
\label{eq:sol-Ak2}
A_k^{-}(\e,\zeta,z) & = & 
\bar a^k 
e^{2ik\e(z-\ln  b/L)/\lambda}
\theta\left(\frac{z-\ln  b/L}{\lambda}-\zeta\right)
\nonumber
\\
& + &
e^{2ik\e\zeta}
\theta\left(\zeta-\frac{z-\ln b/L}{\lambda}\right).
\end{eqnarray}
\end{subequations}

The density of states is described by a solution of Eilenberger equation, represented in the form of series in $a_\pm^k(\e,\x)$. According to Eq.~(\ref{ak-average}), the quantum diffraction leads to a finite separation between points $\x_i$ in the expression for $a_\pm^k(\e,\x)$. The separation is determined by $\rho_{\rm min}$ and $\phi_{\rm min}$ defined by Eq.~(\ref{eq:minpack}).
We conclude, that the Green function may be represented as a sum of functions $A_k^{\pm}(\e,\zeta,\ln \rho_{\rm min}/L)$ over all positive integers $k$, see Eqs~(\ref{ak-average}) and (\ref{eq:Ak-def}). 

The functions 
%\begin{widetext}
\begin{subequations}
\label{78}
\begin{eqnarray}
\label{78a}
\apq(\e,\zeta) & = & 
\left\{%
\begin{array}{ll}
    \exp(2i\e(\zeta-l)), & \hbox{$l-\tte <\zeta<l$;} \\
    a_0, & \hbox{ $0<x<l-\tte $,} \\
\end{array}%
\right.    
\\
\label{78b}
\amq(\e,\zeta) & = & 
\left\{%
\begin{array}{ll}
    \exp(-2i\e\zeta), & \hbox{$0<\zeta<\tte $;} \\
    a_0, & \hbox{$ \tte < \zeta < l $.} \\
\end{array}%
\right.    
\end{eqnarray}
\end{subequations}
%\end{widetext}
allow us to rewrite the sum for $\g(\e,\zeta)$ in terms of $A_k^{\pm}(\e,\zeta,z)$ in a more compact form of Eq.~(\ref{13}).
We introduced a new notation $a_0$ for the zero mode, related to $\bar a$ by the following equation $a_0=e^{-2i\e\tte} \bar a$.
The functions $a_{\pm}(\e,\x)$ have all properties we discussed above. They satisfy the boundary conditions at one terminal of a classical trajectory see Table~\ref{tab:1}, 
and coincide with semiclassical solutions of the Eilenberger equation at small distance $\zeta\lesssim \tau_{\rm E}$ or $l-\zeta\lesssim \tte$from this terminal,
when the quantum diffraction effects may be disregarded. As an electron travels for time interval equal to the Ehrenfest time, 
$\tte$, the quantum scattering suppresses the oscillations and $a_\pm(\e,\zeta)$ jumps to a zero mode value $a_0$.

Using Eqs.~(\ref{78}) to calculate the density of states at large energy, $\e\gg\gesc$, we obtain the result of Eq.~(\ref{N-largeE-QC}). Indeed, by direct averaging of $a_\pm(\e,\x)$ in the form given by Eqs.~(\ref{78}), 
and comparing the result with Eq.~(\ref{69}), we find $a_0=e^{-2i\e\tte}\gesc/(2i\e)$, and then again integrating $a^k_\pm(\e,\x)$ over the phase space we reproduce Eq.~(\ref{77k}), apart of the term $\lambda_2\e^2\tte/\lambda^2$, which is important only at the extremely large energy $\e\gtrsim \sqrt{\lambda/\tte}$.

Below we perform the self-consistent calculations of the Green function in the Riccati parameterization, Eq.~(\ref{13}), with functions $a_\pm(\e,\x)$ 
given by Eqs.~(\ref{78}). Within the anzats of Eqs.~(\ref{78}), there exists only one parameter $a_0$, which completely determines the Green function in the grain. This single parameter can be found with the help of Eq.~(\ref{33}).
Indeed, we substitute $g(\e,\x)$ and $f(\e,\x)$ in terms of the Riccati variables $a_{\pm}(\e,\x)$, write $a_0=\exp(-2i\psi)$ and obtain the following equation for $\psi$:
\begin{widetext}
\begin{eqnarray}
\tan\psi & = & 
\frac{\e}{\gamma_{\rm esc}}
\left(
\frac{e^{-\gamma_{\rm esc}\tte }}{\cos 2\psi}
+2(1-e^{-\gamma_{\rm esc}\tte })\cos 2\psi
\right) + 
2e^{-\gamma_{\rm esc}\tte } \sin 2\psi \ln\frac{\cos\psi}{\cos(\psi-\e\tte)}
\nonumber
\\
& + &
\int_{\tte}^{2\tte}
\gamma_{\rm esc} e^{-\gamma_{\rm esc} (l-\tte)}
\left\{
\frac{\sin \e (2\tte-l)}{\cos\e l}
+2\sin 2\psi \ln\frac{\cos\psi}{\cos(\psi+\e(l-\tte))}
\right\}dl
\label{81}
\end{eqnarray}
Representing $\g(\e,\x)$ in terms of functions $a_{\pm}(\e,\zeta)$, we find
the density of states from Eq.~(\ref{8}):
\begin{eqnarray}
\N(\e) & = & \frac{2}{\mls}
\Biggl[
\frac{\pi^2\gesc^2}{\e^2}
\left(
\sum_{n=0}^{N^*_1}(n+1/2)e^{-\pi \gesc (n+1/2)/\e}+
\sum_{n=N^*_1+1}^{N^*_2}\left(2\tte\e/\pi+(n+1/2)\right)
e^{-\pi \gesc (n+1/2)/\e}
\right)
%\right.
\nonumber
\\
&+&
%\left.
e^{-2\gesc\tte}{\rm Im}\left\{
\tan 2\psi+ \ln \frac{\cos(\psi-\e\tte)}{\cos(\psi+\e\tte)}\right\}
+
\int_{\tte}^{2\tte}\gesc e^{-\gesc l}
{\rm Im}\left\{
\ln \frac{\cos(\psi-\e(l-\tte))}{\cos(\psi+\e(l-\tte))}
\right\}
dl
\Biggr].
\label{82}
\end{eqnarray}
\end{widetext}
Here $N^*_j={\rm int}(j\e\tte/\pi-1/2)$ is the integer part of $j\e\tte/\pi-1/2$. The terms in the first line of Eq.~(\ref{82}) originate from purely ballistic contributions of the electron Green function, while the term, containing $\tanh 2\psi$, represents the contribution of the zero mode. The remaining terms describe the contribution from the parts of the phase space, where both the zero mode and ballistic solutions for the electron-hole wave pairs are intermixed. In the semiclassical limit only the first term in the first line of Eq.~(\ref{82}) survives with $N^*_1\to\infty$, and thus coincides with Eq.~(\ref{i1}). In the quantum limit, $\tte\to 0$, only the  term containing $\tanh 2\psi$ remains in Eq.~(\ref{82}), and corresponds to the random matrix result of Eq.~(\ref{38.7}).

According  to Eq.~(\ref{82}), the density of states is always finite at $\e>\pi/4\tte$, and is formed by at least the second term of the first line of Eq.~(82). We investigate if the gap energy $E_{\rm g}$
is smaller than $\pi/4\tte$, so that the density of states originates from electron states described by the zero mode. 

From Eq.~(\ref{82}) we conclude that the density of states also exists below $\pi/4\tte$ as long as the solution to Eq.~(\ref{81}) has an imaginary part.
For small energies $\e$, Eq.~(\ref{81}) has real solutions and $\N(\e)=0$. 
As energy $\e$ increases, two solutions  approach each other and
at certain value of energy $\e=E_{\rm g}$, they are both equal to some value $\psi_{\rm c}$. The energy $E_{\rm g}$ is the gap energy and the density of states has a square root singularity just above the gap. 
Indeed, we can expand both sides of Eq.~(\ref{81}) to the lowest order in  
$\psi-\psi_{\rm c}$ and $\e-E_{\rm g}$. The expansion in $\e-E_{\rm g}$ linear, but expansion in $\psi-\psi_{\rm c}$ is quadratic, since the first order derivative of Eq.~(\ref{81}) with respect to $\psi$ vanishes at 
$\psi_{\rm c}$, see Sec.~\ref{sec:3.2.1} for a similar analysis.

\begin{figure}
\epsfxsize= 7.5 cm  
\centerline{\epsfbox{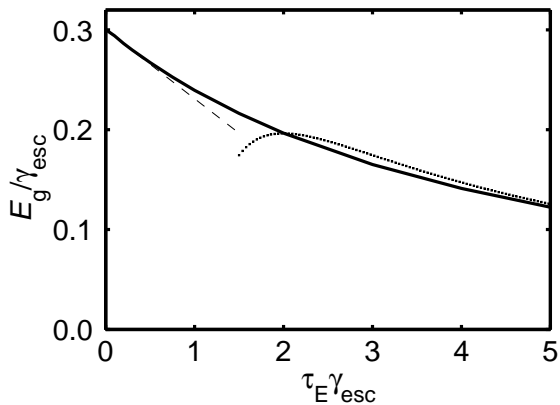}} 
\caption {
The gap energy $E_{\rm g}$ as a function of the Ehrenfest time $\tte$ is shown by a solid line. The dotted line represents the asymptote result for long Ehrenfest time, $\tte\gesc\gg 1$, given by Eq.~(\ref{eq:Eg-asymp}) and the dashed straight corresponds to the short Ehrenfest time limit,  $\tte\gesc\ll 1$, see Eq.~(\ref{83}).
}
\label{fig:10}
\end{figure}

Figure \ref{fig:10} demonstrates dependence of the energy gap $E_{\rm g}$ on the Ehrenfest time $\tte$, shown by a solid line. 
To the lowest order in $\gamma_{\rm esc} \tau_{\rm E}$ the gap energy 
decreases linearly from the random matrix result,\cite{MBFB} see also Eqs.~(\ref{eq:rmt}) and (\ref{41}):
\begin{equation}
\label{83}
E_{\rm g}\approx E_{\rm g}^{\rm rmt}(1-0.23 \gamma_{\rm esc} \tau_{\rm E}).
\end{equation}
The density of states above $E_{\rm g}$ is given by
\begin{equation}
\label{84}
\N(\e)=\frac{c}{\mls}
\sqrt{\frac{\e}{E_{\rm g}}-1},
\end{equation}
where $c\approx c_{\rm rmt}(1-  0.42 \gamma_{\rm esc} \tau_{\rm E})$.
As the Ehrenfest time increases, the energy gap decreases and is given by the following asymptote at $\tte\gesc\gg 1$:
\begin{equation}
E_{\rm g}\approx \frac{\pi}{4\tte}\left(1-\frac{1}{\tte\gesc}\right).
\label{eq:Eg-asymp}
\end{equation}
We plot the asymptote of Eq.~(\ref{eq:Eg-asymp}) by a dotted line in Fig.~\ref{fig:10}.

%%%%%%%%%%%%%%%%%%%%%%%%%%%%%%%%%%%%%%%%%%%%%%%%%%%%%%%%%%%%%%%%%%%%%%%%
%%%%%%%%%%%%%%%%%%%%%%%%%%%%%%%%%%%%%%%%%%%%%%%%%%%%%%%%%%%%%%%%%%%%%%%%
%%%%%%%%%%%%%%%%%%%%%%%%%%%%%%%%%%%%%%%%%%%%%%%%%%%%%%%%%%%%%%%%%%%%%%%%

\section{Discussion and Conclusions}
\label{sec:5}

%%%%%%%%%%%%%%%%%%%%%%%%%%%%%%%%%%%%%%%%%%%%%%%%%%%%%%%%%%%%%%%%%%%%%%%%
%%%%%%%%%%%%%%%%%%%%%%%%%%%%%%%%%%%%%%%%%%%%%%%%%%%%%%%%%%%%%%%%%%%%%%%%

In this paper we considered the effect of quantum disorder and quantum chaos
on electron density of states in an Andreev billiard - a metal grain or a semiconductor dot of irregular shape, connected to a superconductor.

First we restate our assumptions about the considered system and briefly review prior publications, discussing related systems.
We study the density of states at energies much smaller than the gap energy in the superconductor and  the Thouless energy $E_{\rm Th}$ (or the inverse ergodic time $\tau_{\rm erg}$). The opposite limit, when the escape time is equal or less than  the ergodic time, was considered in refs.~[\onlinecite{not0d}] for Andreev billiard, in refs.~[\onlinecite{PBB,OSF,1dU}] for one dimensional systems (wires)
and in refs.~[\onlinecite{SLL,LO}] for bulk superconductors.

In our calculations we use the exponential distribution $P(l)$ of trajectory lengths $l$, described by Eq.~(\ref{50.5}).  An analysis of motion in classically chaotic systems shows deviation of the distribution function from the exponential law due to anomalously long trajectories.\cite{ZE} 
The effect of these trajectories on the density of states of an Andreev billiard was studied in ref.~[\onlinecite{IR}] within semiclassical theory, when the contribution of long trajectories is especially important. 
We assume  that the deviation from the exponential distribution occurs at lengths $l$ longer  than either the Ehrenfest time or the mean free time, and we neglected the contribution of anomalously long trajectories to the electron density of states. 

We study only non-integrable systems. The difference between integrable and chaotic non-integrable systems was examined for a quantum dot in refs.~[\onlinecite{MBFB,CBA}].

We do not consider the effect of finite mean level spacing in the grain.
The finite mean level spacing leads to mesoscopic fluctuations of the energy levels, and particularly, to the fluctuations of the gap.\cite{VBAB} 
The fluctuations of the energy gap smear the square
root singularity of the density of states, Eq.~(\ref{i3}), and produce 
an exponentially small tail of the ensemble average density of states below the gap energy, see refs.~[\onlinecite{BNA,VBAB,OSF,OSF1}]. 
Nevertheless,   the corresponding scale for the gap fluctuations and the mean level spacing near the gap are determined by the following small energy scale: $(\mls^2 \gamma_{\rm esc})^{1/3}\ll \gesc$. Significant fluctuations of the first energy level (energy gap) occur in exponentially rare cases.\cite{BNA,VBAB,OSF,OSF1}

We assume that the reflection at the contact with the superconductor is described by the Andreev mechanism, {\it i.e.} the contact has no tunnel barrier between the grain and the superconductor. The effect of finite barrier was studied in ref.~[\onlinecite{MBFB}] in the random matrix limit,
where it was shown that a weak (normal) reflection at the barrier does not qualitatively change properties of the curve of the density of states.

We do not take into account the Coulomb interaction between electrons in the grain responsible for the effects of zero bias anomaly and the Coulomb blockade. 
In our system, the conductance of the SN contact is large (the number of channels $N\gg 1$), and the Coulomb interaction may be disregarded. The influence of the zero bias anomaly on proximity effect was studied in ref.~[\onlinecite{OBSA}], and the Coulomb blockade in a small metal grain connected to a superconductor  was studied in ref.~[\onlinecite{MatGlaz}].

we also assume that the repulsion in the Cooper channel is sufficiently small and we disregard it. The effect of repulsion was studied, for example in refs.~[\onlinecite{Rep1,Rep2,Rep3}].

The model of clean chaotic Andreev billiards, considered in Sec.~\ref{sec:4} of our paper, was also recently  addressed in paper~[\onlinecite{Bnew}]. 
According to ref.~[\onlinecite{Bnew}], the 
energy gap and the density of states above the gap are determined by the random matrix theory result\cite{MBFB} with the appropriately modified number of channels and the mean level spacing. Although the results\cite{Bnew} qualitatively coincide with ours, they are quantitatively differ from our results, presented in Eqs.~(\ref{83})-(\ref{eq:Eg-asymp}) and in Fig.~\ref{fig:10}. We believe that the calculations\cite{MBFB} within random matrix theory are justified only as $\tte\to 0$, since these calculations take into account only the zero mode. 
In general, as we demonstrated in Sec.~\ref{sec:4}, the ballistic parts of trajectories at length smaller than $\tte$, also known as Hikami boxes, give rise  to a substantial contribution to both the density of states and the self-consistency equation. Consequently, Eqs.~(\ref{81}) and (\ref{82}) cannot be reduced by a change of mean level spacing and the number of channels to the random matrix theory equations of ref.~[\onlinecite{MBFB}], see also Eqs.~(\ref{38.5}) and (\ref{38.7}) in Sec.~\ref{sec:3.2.1}, and thus the quantitative result of ref.~[\onlinecite{Bnew}]
is questionable.

The most interesting qualitative result of our work is the oscillations of the density of states as a function of $\e\tte$. [An oscillating character of the density of states near the gap was also found in ref.~[\onlinecite{AB}].] These oscillations are related to the contribution to $\N(\e)$ from the ballistic part of the electron Green function. Although the full length of different trajectories significantly varies, the Ehrenfest time is nearly the same  for all long trajectories with length $l>\tte$. Indeed, even strong fluctuations of the Lyapunov exponent are averaged out on time scale of the ergodic time $\tau_{\rm erg}$, resulting in weak fluctuations of the Ehrenfest time. The fluctuations of the Ehrenfest time are important only at parametrically large energies $\e\propto \sqrt{\lambda/\tte}$ and lead to suppression of the oscillations. 
Similar oscillations also exist in frequency dependence of the weak localization correction to the conductivity\cite{AL1} and the correlation function of electron energy levels.\cite{AL2} 

We notice that the oscillations discussed above should not be associated with the oscillations of the energy levels in the Wigner-Dyson statistics. The latter oscillations have a much smaller period (equal to the mean level spacing) than  $\tte^{-1}$. The energy level oscillations  were studied in systems with quantum disorder\cite{Efetov}
and with quantum chaos.\cite{Alt,BK}  Similar oscillations are also present in the density of states near the gap with the period 
$\propto \delta_1^{2/3}\gesc^{1/3}$.\cite{VBAB,OSF} In the present paper we do not take into account these oscillations, as well as other mesoscopic effects.

The observation of oscillations 
in the density of states with period $\propto \tte^{-1}$ is the major distinguishing characteristic of quantum chaotic systems  from disordered systems, since in quantum disorder such oscillations are absent. The period of the oscillations may be used to estimate the Ehrenfest time and the Lyapunov exponent in each particular metal grain.

In conclusion, our results allow us to describe the classical-to-quantum  crossover in the density of states of Andreev billiards. Methods, developed in this paper should also be applicable to study this crossover in many other problems of solid state physics and optics.

\section*{Acknowledgements}

We are grateful to I. Aleiner for many fruitful discussions. We also thank P. Brouwer for important suggestions and I. Adagideli and C. Beenakker for their comments.  The work was supported by NSF grant \# 0120702.

\appendix

\section{}
In this appendix we show how the small angle scattering term, which mimics the effect of quantum diffraction, may be used to obtain the results of Sec.~\ref{sec:4.2}.

A small angle scattering potential, with a typical scattering
angle  $\phi_{\rm min}\sim \sqrt{\lambda_{\rm F}/L}$, 
leads to the following form of the self energy part of the Eilenberger equation:
\begin{eqnarray}
\hat \Sigma(\e,\rbf,\n) & = & \frac{1}{\tau_{\rm 0}}\g(\e,\rbf,\n)+
\frac{1}{2\tau_{\rm q}} \nabla_{\n}^2\g(\e,\rbf,\n),
\label{60a}
\end{eqnarray}
where we introduced the full mean free path $\tau_0$,
\begin{subequations}
\begin{eqnarray}
\label{60}
\frac{1}{\tau_{0}} & = & 2\pi\nu n_{\rm i}
\int|V(\n-\n')|^2\ \frac{d\n}{\Omega_d},
\end{eqnarray}
and the transport time $\tau_{\rm q}$,
\begin{eqnarray}
\label{60b}
\!\!\!\!\!\!\!\!\!
\frac{1}{\tau_{\rm q}} & = & 2\pi\nu n_{\rm i}
\int|V(\n-\n')|^2\ |\n\times\n'|^2 \frac{d\n}{\Omega_d}.
\end{eqnarray}
\end{subequations}
The first term in the right hand side of Eq.~(\ref{60a}) does not contribute to the scattering term of the Eilenberger equation. The second term represents small angle diffraction and the Eilenberger equation acquires the form with the  Perron-Frobenius differential operator.
For the two dimensional electron system
$\nabla_{\n}=\partial /\partial \phi$, where $\phi$ is the
angle in the direction of $\n$. we estimate $\tau_{\rm q}$ and $\tau_0$ below.

We use the Ricatti parameterization of the Green function, see Eq.~(\ref{13}). For the scattering term of the Eilenberger equation in the form of Eq.~(\ref{60a}), we rewrite Eqs.~(\ref{14}) as:
\begin{subequations}
\label{14QC}
\begin{eqnarray}
\left(2i\e-{\cal L} \right) \apq  & = &\frac{1}{\tau_{\rm q}} 
\left( \nabla_{\n}^2 \apq-
\frac{2\amq\ [\nabla_{\n} \apq]^2}{1+\apq\amq} 
\right)
,
\label{14aQC}
\\
\left(2i\e+{\cal L} \right) \amq  & = & \frac{1}{\tau_{\rm q}} 
\left( \nabla_{\n}^2 \amq-
\frac{2\apq\ [\nabla_{\n} \amq]^2}{1+\apq\amq} 
\right)
\label{14bQC}
\end{eqnarray}
\end{subequations}
Here again  we omitted variables  $\{\e,\x\}$ of the functions $a_{\pm}=a_{\pm}(\e,\x)$.

We notice that the average of the electron-hole wave function over the phase space is small at large energy $\e\gg\gesc$, see Eq.~(\ref{69}). This observation allows us to neglect non-linear terms in Eqs.~(\ref{14QC}), since these terms would contribute to the higher than the second order in $\gesc/\e$.
We write down equation for $A_2^{\pm}(\e,\x_1,\x_2)$ using the linearized equations of motion for $a_{\pm}(\e,\x)$, Eqs.~(\ref{14QC}). 
\begin{eqnarray}
\label{64withT}
\!\!\!
\left[
4i\e\mp\! {\cal L}_{\rm cm}\!-\!{\cal L}(\rho,\phi)\!
+\!\frac{1}{2\tau_{\rm q}}
\nabla^2_{\phi}
\right]\!\!\!
A_2^{\pm}(\e,\R,{\bf N},\rho,\phi)=0,
\end{eqnarray}
where ${\cal L}_{\rm cm}$ and ${\cal L}(\rho,\phi)$ are defined by Eqs.~(\ref{65}).
As we already mentioned, the non-linear terms in the right hand side of Eqs.~(\ref{14QC}) lead to higher terms in $\gesc/\e$.

After averaging over the position of the center of mass in the phase space, we obtain a counterpart of Eq.~(\ref{Eqs-for-A2}):
\begin{equation}
\label{67withT}
\left(4i\e+s_2\gesc - {\cal L}_{\rho}+
\frac{1}{2\tau_{\rm q}}\nabla^2_{\phi}\right)
A_2^{\pm}(\e,\rho,\phi)
=\gesc.
\end{equation}
Below we assume that $s_2\approx 1$. Introducing new variables $z$ and $\chi$, according to Eq.~(\ref{72}), and then averaging over $\chi$, we arrive\cite{AL1} to the following equation for $A_2^\pm(\omega,z)$ with $\omega=4\e-i\gesc$:
\begin{equation}
\left[
i\omega+\lambda\partial_z +
\frac{\lambda_2}{2}\partial_z^2+
\frac{e^{-2z}}{2\tau_{\rm q}}
\left(
\frac{1-\gamma}{2}\partial_z^2+\gamma\partial_z
\right)
\right]A_2^\pm=\gesc
\label{eq:A2-z-angle}
\end{equation}
where $\partial_z=\partial /\partial z$, and $\gamma$ is a numerical coefficient of order of unity.\cite{AL1}  
Equation~(\ref{eq:A2-z-angle}) should be supplied with the boundary conditions at $z=\ln b/L$, determined by Eq.~(\ref{68}).
The solution can be found as a superposition of two terms, see Eq.~(\ref{70}). One term, $A_{\rm 2,b}^{\pm}(\e,z)$, is given by Eq.~(\ref{71}) and the other term is given by
\begin{eqnarray}
A_{\rm 2,q}^{\pm}(\e,z)& = \!\!&\!\!
\left[
\overline{a_{\pm}(\e,\x)}^2-
A_{2,b}^{\pm}(\e)
\right]
 \frac{w(\omega,z)}{w(\omega,\ln b/L)}.
\label{72.5q}
\end{eqnarray}
Function $w(\omega,z)$ is a solution\cite{AL1} of homogeneous Eq.~(\ref{eq:A2-z-angle}):
\begin{equation}
w(\omega,z)=\exp\left[\left(
\frac{i\omega}{2\lambda}+\frac{\lambda_2\omega^2}{4\lambda^3}
\right)\ln\frac{\lambda\tau_{\rm q}}{\lambda\tau_{\rm q}e^{2z}+\gamma/2}
\right].
\label{75}
\end{equation}

Now solutions for $A_2^{\pm}(\e,z)$ are regular functions as  $z\to -\infty$, and correspondingly the product of two electron-hole pairs, see Eq.~(\ref{A2}), can be found at coincident points.

An exact form and strength of small angle scattering potential is
not important,  since as we will show later, the density of states
depends on the strength of the potential only as a logarithm of
the ratio of the transport time of this potential and the size of
the SN contact. We will choose the strength of the scattering
potential so that at small time this potential scattering would
describe the quantum mechanical diffraction effect. The scattering time $\tau_{\rm q}$ may be estimated as 
$\tau_{\rm q}=\tau_0\phi_{\rm min}^2$, compare  Eqs.~(\ref{60}) and (\ref{60b}). 
We assume that $\lambda\tau_0\sim 1$ (since both $\lambda$ and $\tau_0$ are characterized by the same potential energy of electrons in the dot) and then we estimate $\lambda\tau_{\rm q}\sim \phi_{\rm min}^{-2}=L/\lambda_{\rm F}$, see Eqs.~(\ref{60}) and (\ref{60b}). Combining Eqs.~(\ref{68}) and (\ref{72.5q}) and we obtain Eq.~(\ref{77}).

In conclusion, we discuss the meaning of Eq.~(\ref{75}). Within logarithmic accuracy, the term $\lambda\tau_{\rm q}e^{2z}+\gamma/2$ can be replaced by
${\rm max}\{\lambda\tau_{\rm q}e^{2z},\gamma/2\}$. At large 
scales $\rho$ and $\phi$ of a wave packet, $z>z_{\rm c}=\ln\sqrt{2\lambda\tau_{\rm q}/\gamma}$, we can disregard the quantum diffraction, described by the last term in Eq.~(\ref{64withT}). In this ``classical" approximation we find 
\begin{equation}
w(\omega,z_{\rm c})=\exp\left[\left(
\frac{i\omega}{2\lambda}+\frac{\lambda_2\omega^2}{4\lambda^3}
\right)\ln\frac{2\lambda\tau_{\rm q}}{\gamma}
\right].
\label{eq:w-cl}
\end{equation}
At smaller scales of the wave packet, $z<z_{\rm c}$ or $\rho^2+L^2\phi^2\lesssim \rho_{\rm min}^2$, the quantum diffraction overpowers the Lyapunov classical divergence of trajectories. Consequently, the function $w(\e,z)$ weakly depends on the scale $z$, see Eq.~(\ref{75}), Eq.~(\ref{75}), and we use $w(\varepsilon,z\to -\infty)\approx w(\e,z_{\rm c})$. It is exactly this property of the function $w(\e,z)$ was used in Sec.~\ref{sec:4.2} of the present paper, when we used $w(\e,z_{\rm c})$ 

In this Appendix we used linearized Eqs.~(\ref{14QC}), which are justified at high energy $\varepsilon\gg \gesc$. In general, functions $A_2^{\pm}(\varepsilon,z)$ are described by non-linear Eqs.~(\ref{14QC}). Nevertheless,  the right hand side of Eqs.~(\ref{14QC}), which contains non-linear terms, is necessary only to determine the function $w(\e,z)$ at small scale of the wave packet, {\it i.e.} for $z<z_{\rm c}$, and
at already at scale $z\simeq z_{\rm c}$ the right hand side of Eqs.~(\ref{14QC}) does not affect the solution for $w(\e,z)$ within logarithmic accuracy. That is why the non-linear term is not important within logarithmic accuracy.
 
Within the same logarithmic accuracy, we use approximate solutions for $A_{k}^{\pm}(\e,z)$ at $z\simeq z_{\rm c}$ for $k\geq 2$ in Sec.~{\ref{sec:4.q}}.

%\bibitem{LO_b} A. I. Larkin and Yu. N. Ovchinnikov, red book.

%\bibitem{SB} P.G. Silvestrov and C.W.J. Beenakker,
%Phys. Rev. E {\bf 65}, 035208(R) (2002).

%\end{multicols}
\end{document}